# CIRCUMSTELLAR INTERACTION IN SN 1993J

Claes Fransson and Peter Lundqvist
Stockholm Observatory, S-133 36 Saltsjöbaden, Sweden

and

Roger A. Chevalier
Department of Astronomy, University of Virginia
P.O. Box 3818, Charlottesville, VA 22903

# ABSTRACT

The radio and X-ray observations of SN 1993J during the first year can be consistently explained as a result of interaction of the expanding ejecta with a circumstellar medium. The density of the circumstellar gas can be deduced from the free-free absorption of the radio emission and from the X-ray luminosity. During the first two weeks, both sets of observations indicate a mass loss rate of $\sim 4 \times 10^{-5} M_\odot$ yr$^{-1}$ for a wind velocity of 10 km s$^{-1}$. The subsequent radio and X-ray observations indicate a density gradient $\propto r^{-s}$, with $1.5 \lesssim s \lesssim 1.7$, as opposed to the $r^{-2}$ gradient expected for a steady, spherically symmetric wind. This may either be caused by a variation of the mass loss rate from the progenitor system, or by a non-spherically symmetric geometry. To explain the properties of the X-ray emission, a steep density gradient in the ejecta is needed. During the first months most of the observed X-ray emission originates from the circumstellar shock, which is adiabatic, while the reverse shock is radiative. To avoid excessive Comptonization in the X-ray range collisionless heating must be ineffective. The soft X-rays observed at 220 days probably originate from the reverse shock. The ionization and temperature structures of the circumstellar gas are calculated; we find that the temperature is in excess of $10^5$ K and the medium is nearly completely ionized by the shock radiation after the formation of the shocks. Preacceleration of the circumstellar gas by the radiation from the outbreak can explain the observed high velocity for the circumstellar N V and H$\alpha$ lines. The high luminosity of the lines indicate that the circumstellar medium close to the supernova progenitor had a complex structure.

*Subject headings:* circumstellar matter – $\gamma$-rays: general – radio: general – supernovae: individual (SN 1993J) – X-rays: general

# 1. INTRODUCTION

SN 1993J was discovered on March 28.86 1993 (Ripero & Garcia 1993). The first detection was on March 28.30 at magnitude 13.4, while a photograph taken on March 25.6 showed no supernova brighter than 17 magnitude (van Driel 1993). The explosion date is thus confined to between March 25.6 and March 28.3. Based on the apparent rapid rise around March 28.0 and the roughly 0.2 day for the shock wave to traverse the star (T. Shigeyama, private communication), we adopt March 27.8 as the time of the explosion. There is evidence for the progenitor object (Perelmuter 1993), but its exact nature is still uncertain. Aldering, Humphreys & Richmond (1994) have suggested that the photometry is consistent with two supergiants of spectral classes early B and early K. Based on the early light curve, Nomoto *et al.* (1993), Podsiadlowski *et al.* (1993), Ray *et al.* (1993), Bartunov *et al.* (1994), Utrobin (1994) and Woosley *et al.* (1994) interpret the progenitor as a star of initial mass $\sim 15\ M_\odot$, which has lost all but $0.1 - 0.9\ M_\odot$ of its hydrogen envelope. They propose that Roche lobe overflow in a binary system is responsible for the loss of the hydrogen envelope. A substantial fraction of this mass may have been lost from the system. An alternative scenario based on a single massive star with an initial mass of 30 $M_\odot$, which has lost most of its hydrogen envelope in the form of a wind, has been proposed by Höflich, Langer & Duschinger (1993). In both scenarios one expects a substantial circumstellar density.

Direct evidence for the presence of circumstellar gas comes from observations with IUE and optical telescopes. In particular, IUE observed a strong N V line at $1238 - 1242$ Å (Sonneborn *et al.* 1993; Fransson & Sonneborn 1994) with a width extending up to $\sim 1000$ km s$^{-1}$. From the first observations a few days after the explosion, the flux of the line decreased by a factor $\sim 100$ on a time scale of about a week. The width implies that it must arise outside the ejecta. The high degree of ionization and fast decay constrain it to be formed close to the supernova. This is in contrast to SN 1987A where the emission increased on a time scale of $\sim 400$ days as a result of a light echo in the ring (Lundqvist & Fransson 1991). In the optical, narrow lines of H$\alpha$, He II $\lambda$ 4686, and possibly [Fe X] $\lambda$ 6374, [Fe XI] $\lambda$ 7892 and [Fe XIV] $\lambda$ 5301 were observed (Filippenko & Matheson 1993a; Benetti *et al.* 1994; Cumming *et al.* 1994). The H$\alpha$ line was resolved with a velocity similar to that of the N V line (Cumming *et al.* 1994). The Fe lines had a considerably smaller FWHM of only $\sim 50$ km s$^{-1}$. On April 13, the narrow lines were no longer observable (Filippenko & Matheson 1993b; Cumming *et al.* 1994).

X-ray emission was first detected by ROSAT on April 3.4 in the $0.1 - 2.4$ keV band (Zimmermann *et al.* 1994) and on April 5.25 by ASCA in the $1 - 10$ keV band (Tanaka *et al.* 1993). OSSE on the Compton Observatory detected the supernova at $\gtrsim 100$ keV (Leising *et al.* 1994). The early appearance suggests that the emission has a circumstellar origin, in contrast to the Comptonized emission from the down scattered $\gamma$-rays from the radioactive decay.

Radio emission has been observed with a first possible detection by the VLA at 1.3 cm on April 2.30. Radio turn-ons have since been observed at subsequently longer wavelengths, consistent with free-free absorption in circumstellar gas (Weiler *et al.* 1993). On April 5.7 it was clearly seen with the Ryle telescope at 2 cm with a flux of 3 mJy; no emission was seen the previous day (Pooley & Green 1993). Subsequent observations at this wavelength showed a roughly linear rise up to day 60. The supernova has also been observed at 3 mm (Phillips & Kulkarni 1993a,b).

The presence of radio and X-ray emission together with narrow emission lines is not new for Type II supernovae. For example, both SN 1979C and SN 1980K showed strong radio emission, turning on some time after the optical outburst (Weiler *et al.* 1986). In addition, SN 1980K was detected as an X-ray source by the Einstein Observatory at a luminosity of $\sim 2 \times 10^{39}$ erg s$^{-1}$ (Canizares, Kriss, & Feigelson 1982). All these features have been interpreted as a result of interaction of the supernova ejecta with its circumstellar medium (see Chevalier 1990 for a review). Both these and several other Type II supernovae have been detected several years after the explosion in the optical, long after radioactivity has decayed away, which is consistent with circumstellar interaction (Chevalier & Fransson 1994; hereafter CF94). In this paper we show that the observations of SN 1993J fit into this picture, although there are some deviations from the 'standard' model. Throughout this paper, we take a distance to M81 of 3.6 Mpc (Freedman *et al.* 1994).

## 2. IMPLICATIONS OF THE RADIO LIGHT CURVES

Assuming that the progenitor was a late type supergiant, it is expected to have had a slow moving dense stellar wind. If the mass loss rate is $\dot{M}$, the wind velocity is $v_w$ and the mass loss is spherically symmetric, the number density at radius $r$ is $n_{cs} = \dot{M}/(4\pi\mu m_H v_w r^2)$, where $\mu$ is the mean atomic weight in amu and $m_H$ is the mass of a hydrogen atom. For most of our discussion the density is the relevant quantity, and only the ratio $\dot{M}/v_w$ enters as a parameter. For typical red supergiants the wind velocity is $10 - 20$ km s$^{-1}$ and the mass loss rate is $10^{-6} - 10^{-4}$ $M_\odot$ yr$^{-1}$.

The observed radio turn-ons of previous Type II supernovae have been well described by a model where the turn-on is a result of the decreasing free-free absorption of the ionized stellar wind as the supernova expands (Chevalier 1982b). Although neutral before the outburst, the radiation from the photosphere at the time of shock break-out and from the shock region ionizes and heats the wind to a high temperature (Lundqvist & Fransson 1988). The interaction of the ejecta and stellar wind results in a shock wave propagating into the stellar wind, and a reverse shock into the ejecta.

Hydrodynamical calculations and analytical arguments show that the ejecta density is well described by a power law in radius (or velocity). A similarity solution for the shock radius can then be found (Chevalier 1982a,b). We write the ejecta density as a function of velocity, $v$, as $\rho_{sn} = \rho_0 (t/t_0)^{-3} (v/v_0)^{-n}$, where $\rho_0$ is the density at velocity $v_0$ and time $t_0$. For compact progenitors, the power law index is expected to be in the range $7 - 12$ (e.g., Chevalier & Soker 1989; CF94). An extended red supergiant may, however, have a considerably steeper gradient. The only two independent quantities are then $\rho_0 t_0^3 v_0^n$ and $\dot{M}/v_w$. If both shocks are radiative, as may be the case shortly after the explosion, a thin shell approximation is applicable. Below we show that a circumstellar density variation different from $\rho_w \propto r^{-2}$ is indicated from the observations. We therefore write

$$\rho_w = \frac{\dot{M}}{4\pi v_w r_o^2} \left(\frac{r_o}{r}\right)^s \qquad (2.1)$$

where $r_o$ is a reference radius, corresponding to the mass loss rate $\dot{M}/v_w$. In the following we use $r_o = 10^{15}$ cm. The maximum velocity of the ejecta, $V = R_s/t$, is then given by

$$V = \left[\frac{4\pi \ (3-s)(4-s) \ \rho_0 \ t_0^3 \ v_0^n \ v_w r_o^{2-s}}{(n-3)(n-4) \ \dot{M}}\right]^{1/(n-s)} t^{-(3-s)/(n-s)}. \qquad (2.2)$$

More general cases differ from this approximation by less than $\sim 20\%$, and are discussed in CF94. The outgoing circumstellar shock has a velocity $v_{cs} = \xi \ V(n-3)/(n-s)$ and the reverse shock $v_{rev} = (3-s)V/(n-s)$, where $\xi$ is the ratio of the forward shock radius to the reverse shock radius. Because $V$ is in principle directly observable as the maximum ejecta velocity, we will use it as a parameter in this paper, rather than $\rho_0$.

The free-free absorption optical depth of the unshocked ionized gas at wavelength $\lambda$ is given by

$$\tau_{ff} \approx 0.46 \; \lambda^2 \left(\frac{\dot{M}_{-5}}{v_{w1}}\right)^2 T_5^{-3/2} V_4^{1-2s} \; \left(\frac{t}{11.57 \text{ days}}\right)^{1-2s} \tag{2.3}$$

where $\dot{M}_{-5}$ is the mass loss rate in units of $10^{-5} \, M_\odot \, \text{yr}^{-1}$, $v_{w1}$ the wind velocity in units of 10 km s$^{-1}$, $T_5$ the wind temperature in $10^5$ K, and $V_4$ the ejecta velocity at the reverse shock in $10^4$ km s$^{-1}$. We have evaluated the Gaunt factor for $\lambda = 2$ cm, a number density ratio $n(\text{He})/n(\text{H}) = 0.10$, and $T = 10^5$ K, and assuming $\xi = 1.3$. The data of Van Dyk $et\ al.$ (1993) on 22.5 April indicate that $\tau = 0.5$ at 2 cm at that time. The shock velocity is not directly measurable from the spectra, but is expected to be somewhat larger than the highest ejecta velocity. The blue edge of the H$\alpha$ line showed a velocity of $\sim 19000$ km s$^{-1}$ on 1993 April 11 (e.g., Lewis $et\ al.$ 1994). Earlier spectra show evidence for even higher velocities. The shock front is more directly observable at radio wavelengths using VLBI techniques. The source was resolved by Marcaide $et\ al.$ (1994) on 1 May, yielding a velocity of $(2.4\pm0.15)\times10^4$ km s$^{-1}$ for a uniform disk model, and by Bartel $et\ al.$ (1994) on 27 April, 17 May, and 27 June, 1993, yielding velocities in the range $18000 - 19000$ km s$^{-1}$ with 5–10% uncertainties. We will use $V = 20000$ km s$^{-1}$ on day 10 for most of the paper. When this is substituted in equation (2.3), and taking limb darkening into account, which reduces the mass loss estimate by $\sim 20\%$, one obtains $\dot{M}_{-5}/v_{w1} \approx 3.8 \; T_5^{3/4}$. The temperature of the unshocked wind close to the shock depends on the properties of both the stellar wind and the expansion rate of the supernova, but is typically $\sim 10^5$ K (see § 4.1). We therefore find that $\dot{M}_{-5}/v_{w1} \approx 4$, within approximately a factor of 2. This is an order of magnitude larger than that determined by Panagia, Van Dyk & Weiler (1993). The reasons for this are probably that Panagia $et\ al.$ use a wind temperature of $10^4$ K and an expansion velocity of $10^4$ km s$^{-1}$.

Using the model by Chevalier (1982b) and including limb darkening effects by the wind, we have modeled the radio light curve at 2 cm obtained by Pooley & Green (1993). We assume that the magnetic field and relativistic particle energy densities scale as the shock ram pressure, $\rho v^2$. In figure 1, we show our best fit for $s = 2$, $n = 30$, spectral index $\gamma = 0.7$ (flux $\propto \nu^{-\gamma}$) and $\tau(2 \text{ cm}) = 0.5$ on 22 April, 1993. The fit is considerably worse than for earlier, "normal" Type II supernovae (e.g., Chevalier 1984; Weiler $et\ al.$ 1986). The differences between the model and the observations are especially large for the rising part of the light curve. We have therefore explored varying some of the parameters with respect to the standard model.

The rising part of the light curve is determined mainly by the exponential decrease of the optical depth with time, $F_\lambda \propto e^{-\tau_\lambda [r(t)]}$, and is not appreciably affected by assumptions about the energy density of the magnetic field and relativistic particles. From equation (2.3) and with $T_e \approx$ constant (§4.1), the linear shape of the light curve therefore indicates a density variation different from the standard $\rho_w \propto r^{-2}$.

The optically thin part of the light curve is given by

$$F_{o\ \lambda} \propto \lambda^{\gamma-2}\ t^{-\left\{3+\gamma-[6+\gamma-\frac{s}{2}(3+\gamma)]\frac{(n-3)}{(n-s)}\right\}}. \qquad (2.4)$$

Type II supernovae typically have $\gamma \approx 0.8$ (e.g., Weiler *et al.* 1994). Therefore, a nearly constant flux may be produced by a radial wind with $\rho_w \propto r^{-1.5}$. We find that good fits are indeed obtained for $\rho_w \propto r^{-1.5}$. This is shown in figure 1 as the dashed line for $n = 30$, $\gamma = 0.7$, $n(\text{He})/n(\text{H}) = 0.1$, $T_5 = 1$, and $\dot{M}_{-5}/v_{w1} = 1.35$ at $10^{15}$ cm. The parameters were chosen to fit observations also at other wavelengths. In figure 2 we show observations by Phillips & Kulkarni (1993a) and Weiler *et al.* (1994) together with the $\rho_w \propto r^{-1.5}$ model.

The epochs of the radio turn-ons at different wavelengths are sensitive to the value of $n$. A lower $n$ than $n = 30$ would cause a larger spread in the turn-on times at the different wavelengths than in figure 2. Large values of $n$ were also obtained for SN 1979C and SN 1980K (Chevalier 1984; Weiler *et al.* 1986). The radio evolution in the standard model depends on the assumption that both the particle density and magnetic energy density scale as the ram pressure. If, e.g., the magnetic field decreases slower than $B^2 \propto \rho v^2$, this affects the turn-on in roughly the same way as a large value of $n$. The departure needed from a simple scaling is a factor of $\sim 2$ in the $B^2$ for the $\rho_w \propto r^{-1.5}$ model over a factor of ten in radius to reduce the value of $n$ in a best fit model from $n = 30$ to $n = 15$, though the fits are not as good as in figures 1 and 2. For lower values of $n$ the fits are not compatible to the observations. A change in the relativistic particle density has roughly the same effect on the radio light curves as a change in the magnetic field.

The observations at 3 mm (Phillips & Kulkarni 1993a,b) show a drop in luminosity after $\sim 100 - 150$ days which is consistent with the $r^{-1.5}$ law extending only out to $\sim (2-3) \times 10^{16}$ cm. At larger radii the observations are consistent with an $r^{-2}$ law. In §5, we discuss further the implications of the fact that an $r^{-1.5}$ wind fits better to the radio data than steady, spherically symmetric mass loss during the first $\sim 100$ days

## 3. X-RAY EMISSION

X-ray emission was first detected by ROSAT on April 3.4 (day 7) with a luminosity of $2.94 \times 10^{39}$ erg s$^{-1}$ in the 0.1 - 2.4 keV band (Zimmermann *et al.* 1994). Subsequent observations showed that the luminosity decreased by $\sim 44\%$ from April 3.4 to May 4. It was also detected on April 5.25 by ASCA in the 1 – 10 keV band with a luminosity of $5 \times 10^{39}$ erg s$^{-1}$ (Tanaka *et al.* 1993). For a thermal bremsstrahlung spectrum the ASCA team found $kT \gtrsim 10$ keV and a column density $\sim 10^{21}$ cm$^{-2}$. The latter is somewhat larger than deduced from the ROSAT observations, $\sim 5.9 \times 10^{20}$ cm$^{-2}$, which were consistent with the galactic absorption in the M 81 direction, $\sim 4.3 \times 10^{20}$ cm$^{-2}$.

Observations using OSSE on the Compton Observatory have revealed SN 1993J as a source of hard photons between $50 - 150$ keV (Leising *et al.* 1994). During the first two observations on days $9 - 15$ and days $23 - 36$ the intensities at 100 keV were $1.82 \times 10^{-3}$ photons cm$^{-2}$ s$^{-1}$ MeV$^{-1}$ and $0.89 \times 10^{-3}$ photons cm$^{-2}$ s$^{-1}$ MeV$^{-1}$, respectively. A third longer observation on days $93 - 121$ gave only an upper limit. With photon spectral indices, $-2.3$ and $-2.2$, respectively, the luminosities between $50 - 150$ keV are $5.5 \times 10^{40}$ erg s$^{-1}$ on day 12 and $3.0 \times 10^{40}$ erg s$^{-1}$ on day 28.5.

The X-ray emission may originate either in the reverse shock or the circumstellar shock. In the thin shell approximation the temperature of the wind shock is

$$T_{cs} = 2.27 \times 10^9 \; \mu_s \; \frac{(n-3)^2}{(n-s)^2} \; V_4^2 \; \text{K}, \tag{3.1}$$

while the reverse shock temperature is

$$T_{rev} = \frac{(3-s)^2}{(n-3)^2} \; T_{cs}, \tag{3.2}$$

assuming electron-ion equipartition in both cases. The mean mass per particle (electrons and ions) in a.m.u. is $\mu_s = [1 + 4 \, n(\text{He})/n(\text{H})]/[2 + 3 \, n(\text{He})/n(\text{H})]$; for solar abundances $\mu_s = 0.61$, while for $n(\text{He})/n(\text{H}) = 1$, as may be the case for SN 1993J (Baron $et\ al.$ 1994), $\mu_s = 1$. If electrons and ions are not in equipartition the ion temperature is a factor $\sim 2$ higher than the above estimates.

The density behind the circumstellar shock is $\rho_{cs} = 4 \, \rho_w$, where $\rho_w$ is given by equation (2.1). The density ratio of the reverse and circumstellar shocks is given by

$$\rho_{rev} = \frac{(n-3)(n-4)}{(3-s)(4-s)} \; \rho_{cs}. \tag{3.3}$$

The swept up mass behind the circumstellar shock is

$$M_{cs} = \frac{\dot{M} R_s}{(3-s) \; v_w} \; \left(\frac{R_s}{r_o}\right)^{(2-s)}, \tag{3.4}$$

and that behind the reverse shock

$$M_{rev} = \frac{(n-4)}{(4-s)} \; M_{cs}. \tag{3.5}$$

For $n = 6 - 30$, the temperature behind the wind shock is $\gtrsim 10^9$ K, while for the reverse shock the temperature is $10^7 - 5 \times 10^8$ K. The time scale for equipartition between electrons and ions is

$$t_{e-i} \approx 2.5 \times 10^6 \left(\frac{T_e}{10^9 \; \text{K}}\right)^{1.5} \left(\frac{n_e}{10^8 \; \text{cm}^{-3}}\right)^{-1} \text{s} \tag{3.6}$$

(Spitzer 1978; Stepney 1983). At 10 days $n_e \lesssim 3 \times 10^8$ cm$^{-3}$ and $T_e \lesssim 3 \times 10^8$ K behind the reverse shock, so $t_{e-i} \lesssim 8.5$ days; the reverse shock is marginally in equipartition, even without collisionless heating. The ion temperature behind the wind shock is $(1-2) \times 10^{10}$ K for $V_4 = 2$, and the density is a factor $\gtrsim 4$ lower than that behind the reverse shock. Electron-ion collisions are therefore highly ineffective, and the electron temperature may be much lower than the ion temperature, unless we invoke efficient plasma instabilities heating the electrons collisionlessly in the shock front.

The higher density behind the reverse shock means that the X-ray emission below 10 keV is dominated by the reverse shock (see below). A possible exception may occur if inverse Compton scattering of photospheric photons by the relativistic electrons in the shocked region is important. However, the observed flat spectrum (Tanaka *et al.* 1993; Zimmermann *et al.* 1994) is incompatible with a spectrum dominated by Comptonization. Below we use this to constrain the electron temperature of the wind shock.

For the reverse shock to dominate the *observed* flux, absorption by the postshock gas cannot be large, as is inferred from the ROSAT observations during the first $\sim 100$ days. A cool, absorbing shell will be present if the reverse shock is radiative. *Therefore, for the observed X-ray flux to come from the reverse shock a necessary condition is that this shock is adiabatic.* At temperatures above $\sim 2 \times 10^7$ K, bremsstrahlung dominates the cooling, and the cooling time behind the reverse shock is $t_{cool} = 1.7 \times 10^{11} \, T_{rev}^{0.5}/n_{rev}$ s, or

$$t_{cool} = 1.6 \times 10^3 \frac{(3-s)^2(4-s)}{2\,(n-s)(n-3)(n-4)} \left(\frac{\dot{M}_{-5}}{v_{w1}}\right)^{-1} V_4^{5-s} \left(\frac{t}{11.57 \text{ days}}\right)^{4-s}$$

$$= C_{n,s} \left(\frac{\dot{M}_{-5}}{v_{w1}}\right)^{-1} V_4^{5-s} \, t_{days}^{4-s} \text{ days.}$$

$$(3.7)$$

In Table 1 we give $C_{n,s}$ for different values of $n$ and $s$, together with $t_{cool}$ for the specific case $\dot{M}_{-5}/v_{w1} = 2$ and $V_4(10 \text{ days}) = 2$. It is seen that $t_{cool}$ decreases rapidly as the density gradient gets steeper. According to equation (2.2) $t_{cool} \propto \dot{M}^{-(15+s^2-9s+n)/(n-s)} \, t^{(4n-ns-15+4s)/(n-s)}$, and cooling is therefore most important at early times. If cooling occurs early on, cool gas is likely to be present between the reverse shock and the contact discontinuity later, giving rise to an X-ray absorption much higher than observed (e.g., CF94). The importance of cooling therefore depends on the time of shock formation. To be conservative we evaluate $t_{cool}$ one day after the explosion in Table 1. For $\dot{M}_{-5}/v_{w1} \gtrsim 2$ and $V_4(10 \text{ days}) \lesssim 2$ the reverse shock becomes radiative for $n \gtrsim 9$. An earlier date would decrease this limit further. Equation (3.7) assumes that only bremsstrahlung is important for the cooling. In §3.2 we discuss the case when $T_e \lesssim 2 \times 10^7$ K, as is the case for $n \lesssim 15$ (eqn. [3.14]).

Summarizing, there are therefore two possible scenarios for the X-ray emission. *Either both shocks are adiabatic, and most flux below $\sim 10$ keV comes from the reverse shock, or alternatively, the flux from the radiative reverse shock is absorbed by cool gas, and all the observed flux originates from the circumstellar shock.* The latter case occurs for a steep ejecta density gradient. We discuss each of these cases separately below.

In either case we can estimate the free-free luminosity from the two shocks from $L_k = 4\pi \int j_{ff}(T_e)(\rho/m_H)^2 r^2 dr \approx j_{ff}(T_k) \, M_k \, \rho_k/m_H^2$, where $k$ denotes either the reverse or circumstellar shock. The emissivity per unit mass is $j_{ff}\rho/m_H^2$ with

$$j_{ff}(E) = 1.64 \times 10^{-20} \ \zeta \ g_{ff} \ T_e^{-0.5} \ e^{-E/kT_e} \ \text{erg cm}^3 \ \text{s}^{-1} \ \text{keV}^{-1}, \qquad (3.8)$$

where $g_{ff}$ is the free-free Gaunt factor, which includes relativistic effects, and $\zeta = [1 + 2 \ n(\text{He})/n(\text{H})]/[1 + 4 \ n(\text{He})/n(\text{H})]$. For $n(\text{He})/n(\text{H}) = 0.1, \zeta = 0.86$, and for $n(\text{He})/n(\text{H}) = 1, \zeta = 0.60$. Because of occultation by the ejecta, only half of the above luminosity escapes outward.

As long as the emission in the ASCA and ROSAT energy ranges comes from the same region the ratio of the ASCA/ROSAT flux is given by

$$\frac{L_{ASCA}}{L_{ROSAT}} = \frac{\int_{1 \text{ keV}}^{10 \text{ keV}} g_{ff}(E) \ e^{-E/kT_e} \ dE}{\int_{0.1 \text{ keV}}^{2.4 \text{ keV}} g_{ff}(E) \ e^{-E/kT_e} \ dE} \approx 2.4 \ e^{-7.9 \times 10^7/T_e} \qquad (3.9)$$

within $\sim 5\%$ for $10^8 \lesssim T_e \lesssim 10^9$ K. The observed ratio is $\gtrsim 1.5$, which sets a lower limit to the electron temperature of $\sim 2 \times 10^8$ K for the X-ray emitting gas.

The density behind the reverse shock is large and the cooling time may become short before the contact discontinuity is reached. Lack of electron-ion equipartition in the circumstellar shock leads to an electron temperature much lower than the ion temperature. Cooling and deviations from equipartition modifies both the density, temperature and the emission from the shock, compared to the adiabatic similarity solution. We have therefore calculated the hydrodynamical evolution of the interaction numerically, using an explicit, Lagrangian code with artificial viscosity. Radiative cooling is included according to Raymond, Cox, & Smith (1976), as well as Compton cooling and heating by both the radiation from the shocked gas and the diluted photospheric flux. Above $\sim 2 \times 10^7$ K cooling of the reverse shock is dominated by free-free emission. We calculate this taking relativistic effects into account, including both electron-ion and electron-electron bremsstrahlung (e.g., Stepney & Guilbert 1983). The case that the electrons are assumed to be heated by Coulomb collisions only (Spitzer 1978; Stepney 1983) is referred to as $T_e = T_{Coul}$. To examine the importance of collisionless heating we have assumed that $T_e = T_{ion}$ in some models.

### 3.1. The adiabatic case

We first discuss the adiabatic case. This may not be relevant for SN 1993J, but is of interest for other supernovae with shallow density gradients of their ejecta. In this case the OSSE flux is dominated by the circumstellar shock, unless the ejecta density gradient is extremely flat, while the reverse shock is responsible for the low energy flux.

The lower limit of 10 keV for the reverse shock temperature at 10 days (Tanaka *et al.* 1993) implies in this case that $n \lesssim [4.4 \ \mu_s^{0.5} \ V_4 \ (3 - s) + s]$. For $V_4 < 2$ we find $n \lesssim 15$. Any departure from electron-ion equipartition decreases this limit further. As we found above, an adiabatic shock requires $n \lesssim 9$, so this constraint is consistent.

At $\sim 1$ keV, $g_{ff} \approx 1.87\, T_8^{0.264}$, where $T_8 = (T_e/10^8\ \mathrm{K})$. The ROSAT luminosity from the reverse shock is therefore

$$L_{rev}(1\ \mathrm{keV}) \approx 1.74 \times 10^{37}\ \zeta\ \frac{4\,(n-3)(n-4)^2}{(3-s)^2(4-s)^2}\ T_8^{-0.24}\ e^{-0.116/T_8}$$
$$\left(\frac{\dot{M}_{-5}}{v_{w1}}\right)^2\ V_4^{3-2s}\ \left(\frac{t}{11.57\ \mathrm{days}}\right)^{3-2s}\quad \mathrm{erg\ s}^{-1}\ \mathrm{keV}^{-1}. \tag{3.10}$$

For a flat spectrum, the ROSAT observation at 7 days corresponds to $1.3 \times 10^{39}\ \mathrm{erg\ s}^{-1}\ \mathrm{keV}^{-1}$. With $T_8 \gtrsim 1$, as indicated by the observations, and $V_4 = 2$, and assuming that the reverse shock is responsible for the X-ray emission, we find $\dot{M}_{-5}/v_{w1} = 2.9$, 1.7, and 1.1 for $n = 6$, 7, and 8, respectively. With $s = 1.5$ the corresponding mass loss rates measured at $10^{15}$ cm are $\dot{M}_{-5}/v_{w1} = 5.0$, 2.9, and 1.9, respectively. The non-uniform density behind the reverse shock tends to decrease the mass loss rate by up to a factor two (see below).

For $T_{rev} \lesssim 5 \times 10^8$ K the reverse shock contributes only marginally to the OSSE luminosity. With $g_{ff}(100\ \mathrm{keV}) \approx 0.32\,(1 + 3\,T_9)$ in the range $0.1 < T_9 < 10$, the luminosity at 100 keV is

$$L_{cs}(100\ \mathrm{keV}) \approx \frac{3.8 \times 10^{36}}{(3-s)}\ \zeta\ T_{e9}^{-0.5}\ (1 + 3\,T_{e9})\ e^{-1.16/T_{e9}}\left(\frac{\dot{M}_{-5}}{v_{w1}}\right)^2$$
$$V_4^{3-2s}\ \left(\frac{t_{days}}{11.57}\right)^{3-2s}\quad \mathrm{erg\ s}^{-1}\ \mathrm{keV}^{-1}. \tag{3.11}$$

At 12 days the observed OSSE flux at 100 keV (Leising *et al.* 1994) corresponded to $4.3 \times 10^{38}\ \mathrm{erg\ s}^{-1}\ \mathrm{keV}^{-1}$. With $V_4 = 2$ the required mass loss rates are $\dot{M}_{-5}/v_{w1} = 14.6$, 9.8, and 8.2 for $s = 2$ and $T_{e9} = 1$, 2, and 3, respectively. With $s = 1.5$ we get $\dot{M}_{-5}/v_{w1} = 12.4$, 8.5, and 7.1, respectively, at $10^{15}$ cm. These are considerably higher than those derived above from the ROSAT and ASCA fluxes using equation (3.10).

The ratio of the OSSE and ROSAT luminosities is given by equations (3.10) and (3.11),

$$\frac{L_{cs}(100\ \mathrm{keV})}{L_{rev}(1\ \mathrm{keV})} = 0.22\ \frac{(3-s)(4-s)^2}{4\,(n-3)(n-4)^2}\ \frac{T_{e\ rev8}^{0.24}\,(1 + 3\,T_{e\ cs9})}{T_{e\ cs9}^{0.5}}\ e^{-1.16/T_{e\ cs9}}. \tag{3.12}$$

The observed ratio is $L(100\ \mathrm{keV})/L(1\ \mathrm{keV}) \approx 0.33$. The theoretical ratio is sensitive to both $n$ and $T_{e\ cs}$. However, for $s = 2$, $n \gtrsim 6$ and $2 \lesssim T_{e\ rev8} \lesssim 5$ the maximum ratio is only 0.11 for $T_{e\ cs} \lesssim 3 \times 10^9$ K. With $s = 1.5$ the maximum ratio is 0.25. The strong dependence on $n$ comes from the ratio of the emission measures of the reverse and circumstellar shocks. *If* both shocks are adiabatic the observations thus favor low values of both $n$ and $s$.

To include cooling and lack of ion-electron equipartition, as well as to include a non-uniform density, we have calculated the shock structure numerically. In the adiabatic case we have varied the power law index from $n = 5$ to $n = 7$. The mass loss rate has been adjusted to give approximate agreement with the ROSAT and ASCA luminosities at $\sim 10$ days. Except where otherwise stated, the ejecta velocity at 10 days has been set at $V_4(10 \text{ days}) \approx 2$. The initial radius, $R_o$, is taken as the radius where the ejecta and wind densities are equal; for $\dot{M}_{-5}/v_{w1} = 2$, $R_o \approx 5 \times 10^{13}$ cm.

In figure 3 we show for $T_e = T_{Coul}$ the electron and ion temperatures and the density in the interaction region for $n = 6$ and $\dot{M}_{-5}/v_{w1} = 3$, at day 10. The density structure is similar to the similarity solution with a dense, shocked ejecta shell and a low density shocked wind. The only difference is close to the contact discontinuity, where the finite initial radius gives a less peaked density. Immediately behind the reverse shock the electron temperature is a factor $\sim 2$ smaller than the ion temperature, but has reached the ion temperature halfway through the shell. The main result of the numerical calculations is that behind the wind shock the electron temperature is much lower than the ion temperature, and $T_e$ does not rise much higher than $\sim 10^9$ K at any time. This has important consequences for the X-ray emission from the wind shock.

For $\dot{M}_{-5}/v_{w1} = 3$ we find cooling to become important for $n \gtrsim 8$. This leads to a catastrophic fall in $T_e$ below $\sim 2 \times 10^7$ K, down to $\sim 5 \times 10^3$ K where photoelectric and Compton heating stabilizes the temperature (CF94). These calculations basically confirm the estimates from equation (3.7), although the exact borderline between the adiabatic and cooling case is constrained to $n \lesssim 7$ for $\dot{M}_{-5}/v_{w1} \gtrsim 2$ and $V_4(10 \text{ days}) = 2$. A higher value of $V$, or lower $\dot{M}$, relaxes this condition (eqn. [3.7]).

In the adiabatic models, most of the observed X-ray emission below $\sim 10$ keV comes from the reverse shock wave, while the forward shock front is a source of hard X-rays ($\gtrsim 100$ keV). The larger emission measure behind the reverse shock causes the emission from this region to dominate up to high energies. For $n \lesssim 5$ the temperature of the reverse shock is comparable to that behind the circumstellar shock, $\sim 10^9$ K, if $T_e = T_{Coul}$.

In Table 2 we give the luminosities at 10 days in the ROSAT, ASCA and OSSE bands for $n = 5 - 7$, $V_4(10 \text{ days}) \approx 2$, $s = 2$, and for different $\dot{M}_{-5}/v_{w1}$. The $n = 7$, $\dot{M}_{-5}/v_{w1} = 1.5$ model gives a fair agreement with the observed fluxes in the ROSAT and ASCA bands. The ASCA/ROSAT ratio is uncomfortably low because of the low temperature in the shocked ejecta, especially close to the contact discontinuity. The analytical estimates from equations $(3.10 - 3.12)$ are accurate to within a factor of $\sim 2$, and these expressions are therefore useful for estimating the luminosities. The most serious objection to this model for the early X-rays is that the OSSE luminosity is far too low, even with collisionless heating included, as shown by the $T_e = T_{ion}$ models. A larger $\dot{M}$ increases the OSSE luminosity, but then overproduces the low energy luminosities. Increasing $V(10 \text{ days})$ also increases the OSSE flux, but even models with $V_4(10 \text{ days}) = 3$ are not sufficient for this $\dot{M}$ and $n$. Models with $n = 6$, or lower, in combination with a larger $\dot{M}$ give an improved fit to the combined X-ray and OSSE observations. The lower $n$ results in a higher reverse shock temperature, and larger $50 - 150$ keV flux, while a higher $\dot{M}$ is needed to compensate for the decrease in emission measure with decreasing $n$ (eqn. [3.10]). Because of the higher ion temperature, electron-ion equipartition is slower, and we find that, unless collisionless heating is important, it is impossible to reproduce the OSSE luminosity for the $n = 6$ model. Retaining the other parameters of the model, equipartition increases the calculated OSSE luminosity by a factor $\sim 3$ compared to the $T_e = T_{Coul}$ model, still a factor $\sim 4$ low compared to the observations. Models with $n = 5$ in Table 2 continue this trend, but these also fail to give a high enough OSSE flux, confirming the analytical discussion.

The spectrum of the model in figure 3 at 10 days is shown in figure 4, together with the $T_e = T_{ion}$ model. The contributions from the reverse and circumstellar shocks are shown separately, and it is clear that the circumstellar contribution is small even in the OSSE range in the pure Coulomb heating case. The circumstellar shock is important only for the $T_e = T_{ion}$ model, and then only above $\sim 100$ keV. In these spectra we have *not* included the Comptonization of the photospheric photons. Because of the very high electron temperature in the $T_e = T_{ion}$ model, $T_e \approx 6.9 \times 10^9$ K, Comptonization dominates the X-ray emission below $\sim 10$ keV in contradiction to observations (§3.2 and Table 4).

If the wind density decreases slower than $r^{-2}$, as indicated by the radio observations during the first $\sim 100$ days, this leads to a slower decline of the X-ray flux. If the reverse shock is adiabatic at the observed epochs it will remain so. The total X-ray luminosity decreases as $t^{-[12-7s+n(2s-3)]/(n-s)}$. The reverse shock temperature is falling as $t^{-[2(3-s)]/(n-s)}$, and the luminosity falling in a given band with $E \ll kT_{rev}$ depends on time as

$$L(E) \propto t^{-[(2s-3)n-5s+6]/(n-s)}. \tag{3.13}$$

For $s = 2$, $n = 7$, $L(E) \propto t^{-0.6}$, while $s = 1.5$ and $n = 7$ give $L(E) \propto t^{0.27}$. Neglecting the weak temperature dependence on the Gaunt factor, the observed decline could in the adiabatic case be reproduced by $s \approx 1.85$. However, we again stress that this case is probably not applicable to SN 1993J.

### 3.2. *The radiative case*

Radiative cooling at the reverse shock front becomes important for $n \gtrsim 8$. Above $2 \times 10^7$ K, the cooling is dominated by bremsstrahlung and the cooling time is given by equation (3.7). Below $2 \times 10^7$ K, the cooling is dominated by lines and the cooling rate is $\Lambda(T_e) \approx 3.4 \times 10^{-23} T_7^{-0.67}$ erg cm$^3$ s$^{-1}$ (Raymond, Cox & Smith 1976). In the radiative case, the cooling time of the reverse shock is therefore

$$t_{cool} = 3.5 \times 10^9 \frac{(4-s)\,(3-s)^{4.34}}{(n-3)(n-4)(n-s)^{3.34}} \; V_4^{3.34+s} \; \left(\frac{\dot{M}_{-5}}{v_{w1}}\right)^{-1} \left(\frac{t}{11.57 \text{ days}}\right)^s \text{ s}. \tag{3.14}$$

The reverse shock is radiative up to at least one year for all $\dot{M}_{-5} \, v_{w1}^{-1} \gtrsim 2$, $V_4 \lesssim 2$ and $n \gtrsim 15$. However, the cooling time is sensitive to $s$ and especially to $n$, $t_{cool} \propto n^{-5.34}$ for $n \gg 4$. A shallower density gradient, as is likely in the interior, increases the cooling time scale considerably.

All X-rays from the reverse shock are absorbed before $\sim 50$ days, and only emission from the circumstellar shock is observed. The cool shell between the reverse shock and the contact surface has a column density

$$N_{cool} = 9.07 \times 10^{21} \, \frac{2 \, (n-4)}{(3-s)(4-s)} \left( \frac{\dot{M}_{-5}}{v_{w1}} \right) \, V_4^{1-s} \, \left( \frac{t}{11.57 \text{ days}} \right)^{1-s} \text{cm}^{-2}. \quad (3.15)$$

X-rays below

$$E_{\tau=1} = 1.23 \left[ \frac{2 \, (n-4) \, \dot{M}_{-5}}{(3-s)(4-s) \, v_{w1}} \right]^{3/8} \, V_4^{-3(s-1)/8} \, \left( \frac{t}{11.57 \text{ days}} \right)^{-3(s-1)/8} \text{keV} \quad (3.16)$$

are absorbed, and the temperature of the reverse shock therefore has to be less than $\sim E_{\tau=1}/k$. At 10 days this is for SN 1993J satisfied for $n \gtrsim 13$ if $V_4 \gtrsim 2$ and $\dot{M}_{-5}/v_{w1} \approx 4$. The late X-ray observations, discussed below, indicate $n \gtrsim 20$ and the cooling constraint is well satisfied.

If cooling, the total energy emitted from the reverse shock is

$$\begin{aligned} L_{rev} &= \frac{(3-s)^2(n-3)(n-4)}{2 \, (4-s)(n-s)^3} \, \frac{\dot{M}V^3}{v_w} \left( \frac{r}{r_o} \right)^{2-s} \\ &= 1.57 \times 10^{41} \, \frac{2 \, (3-s)^2(n-3)(n-4)}{(4-s)(n-s)^3} \, \frac{\dot{M}_{-5}}{v_{w1}} \, V_4^{5-s} \left( \frac{t}{11.57 \text{ days}} \right)^{2-s} \text{erg s}^{-1}. \end{aligned} \quad (3.17)$$

Half of this is emitted outwards. Both the energy absorbed by the ejecta and the cool shell contribute to the bolometric luminosity of the supernova. As long as cooling is important, the luminosity decreases as $L_{rev} \propto t^{-(15-6s+sn-2n)/(n-s)}$. This energy input is mainly deposited in the outer ejecta close to the shock, and in the cool, shocked gas. This may give rise to broad emission lines in the UV and to H$\alpha$ (e.g., Fransson 1984; CF94). A high density in the ejecta and cooling shell at early times may thermalize the lines (Baron *et al.* 1994).

At $10^9$ K $< T_e < 5 \times 10^9$ K the Gaunt factor at 1 keV is $g_{ff} = 3.45 \, T_9^{0.663}$ to within $\sim 2\%$. The ROSAT luminosity is then

$$\begin{aligned} L_{cs}(1 \text{ keV}) \approx \; &\frac{4.1 \times 10^{37}}{(3-s)} \, \zeta \, T_9^{0.16} \, e^{-0.0116/T_9} \\ &\left( \frac{\dot{M}_{-5}}{v_{w1}} \right)^2 \, V_4^{3-2s} \, \left( \frac{t}{11.57 \text{ days}} \right)^{3-2s} \text{erg s}^{-1} \text{ keV}^{-1}. \end{aligned} \quad (3.18)$$

With $V_4 = 2$ and $\zeta = 0.86$ we obtain $\dot{M}_{-5}/v_{w1} \approx 6.7$ ($s = 2$) and $\dot{M}_{-5}/v_{w1} \approx 7.5$ ($s = 1.5$), nearly independent on temperature. The ASCA/ROSAT ratio is 2.3 – 2.6 for $10^9$ K $< T_e < 5 \times 10^9$ K, within the observed range.

Also in this case, the OSSE flux is given by equation (3.11), and the same values for $\dot{M}/v_w$, depending on temperature, are required. The OSSE/ROSAT ratio is just given by the form of the free-free spectrum and can be approximated by

$$\frac{L(100\ \text{keV})}{L(1\ \text{keV})} = \frac{g_{ff}(100\ \text{keV})}{g_{ff}(1\ \text{keV})}\ e^{-1.16/T_{e\ 9}} \approx\ 0.39\ T_{e\ 9}^{0.143}\ e^{-1.16/T_{e\ 9}}. \quad (3.19)$$

For $T_e \gtrsim 1.2 \times 10^9$ K the ratio is within a factor two of the observed one, which is $L(100\ \text{keV})/L(1\ \text{keV}) \approx 0.33$. This ratio is considerably easier to obtain in the cooling scenario.

The importance of cooling, as well as departure from electron-ion equipartition, makes a numerical calculation necessary. For the low reverse shock temperatures in these models the cooling time is very short, with gas cooling from $\sim 10^7$ K down to $\sim 10^4$ K in less than one day at an age of 10 days for $n \gtrsim 15$ (eq. [3.14]). The short cooling time behind the reverse shock makes an accurate calculation of its structure difficult. To tackle this problem we have used two approaches, either by a modified hydrodynamic method, or by using an extension of the similarity solution.

In the hydrodynamic approach we avoid too short time steps by merging the cool shells continously. Even then the structure was difficult to resolve, and the total number of time steps became more than $10^6$. Instead of attempting even finer zoning and smaller time steps, we take advantage of the short cooling time and calculate the structure of the reverse shock assuming a stationary structure, with realistic cooling and emission from the gas. The temperature and density immediately behind the shock are determined from the time dependent solution.

In the other approach we use the similarity solution for an adiabatic circumstellar shock and a radiative reverse shock (CF94). This applies if cooling of the circumstellar shock is negligible and if the finite initial radius is ignored, i.e. after $t \gg R_o/V_s \approx 0.2$ days. The electron-ion heating of the shocked circumstellar gas is calculated by following each mass element, starting at the shock. The structure of the cooling shock is calculated as above. After solving the ordinary differential equations describing the density and temperature behind the reverse shock and including cooling, we calculate the spectrum from the cooling gas. The emission is assumed to be in equilibrium at the local temperature and density. The emissivity is calculated using recombination and collisional ionization rates from Arnaud & Rothenflug (1985), and collisional excitation rates from Gaetz & Salpeter (1983), Mewe & Gronenschild (1981), and Raymond & Smith (1979). To calculate the photoelectric absorption of the cool gas between the shocks we assume this gas to be neutral. The spectrum from the circumstellar shock is treated as for the adiabatic shock.

Figure 5 shows the temperature and density at 10 days for the $s = 2$, $n = 25$, $\dot{M}_{-5}/v_{w1} = 3.5$ and $T_e = T_{Coul}$ model, calculated with the full hydrodynamics. The mass loss rate was adjusted to approximately give the observed ROSAT and ASCA luminosities at 10 days. Both density and temperature are nearly constant behind the circumstellar shock. As with the model in figure 3, the electron temperature is only $\sim 1.6 \times 10^9$ K, much lower than the ion temperature. The cooling shell behind the reverse shock is very thin and can only be seen as a spike in the density distribution and a similar dip in the temperature. In figure 6 we show an expanded view of the structure of the reverse shock, calculated from the stationary solution. The cooling takes place over a very narrow spatial interval, and has basically the structure of a thermal instability. Most of the energy is released in this region. The cool shell is likely to be broken up by instabilities in a multi-dimensional calculation. The column density of this shell corresponds well to that obtained from equation (3.15). The $T_e = T_{ion}$ model is similar, but has an electron temperature $\sim 5 \times 10^9$ K.

In Table 3 the ROSAT and ASCA luminosities are given, as well as the OSSE 100 keV luminosity, at 10 days for different values of $s$, $\dot{M}$ and assumptions about electron-ion equipartition. The value of $n$ is of minor importance at this early epoch, as no emission from the reverse shock penetrates the cool shell. As expected from equation (3.18) the luminosity scales roughly as $\dot{M}^2$. The mass loss rate was adjusted to give approximately the correct ROSAT and ASCA luminosities, and does not vary much with various assumptions of equipartition, $n$ or $s$. We find that $\dot{M}_{-5}/v_{w1} \approx (4-5)$ is required. Compared to the values of $\dot{M}/v_w$ derived from equation (3.18), this is smaller by a factor $\sim 1.7$, which is mainly a result of the nonuniform density structure. As shown by the $T_e = T_{ion}$ and $T_e = T_{Coul}$ models the OSSE flux is sensitive to the degree of electron-ion equipartition. The $T_e = T_{ion}$ model has a flux higher by a factor of $\sim 2$, compared to the $T_e = T_{Coul}$ model. The OSSE/ROSAT ratio is close to the maximum value inferred from equation (3.19). Although somewhat low, even the $T_e = T_{Coul}$ models give acceptable fits to the observed $L(100\ \mathrm{keV}) \approx 4.3 \times 10^{38}$ erg s$^{-1}$ keV$^{-1}$.

A high $T_e$, as in the $T_e = T_{ion}$ models with collisionless heating, increases the Comptonized flux in the keV range considerably. The electron scattering optical depth behind the circumstellar shock is

$$\tau_e = \frac{2.0 \times 10^{-2}}{(3-s)}\ \zeta\ \frac{\dot{M}_{-5}}{v_{w1}}\ V_4^{1-s}\ \left(\frac{t}{11.57\ \mathrm{days}}\right)^{1-s}. \qquad (3.20)$$

In each scattering the photons increase their energy by a factor $\Delta\nu/\nu \approx 4kT_e/m_ec^2 \approx 1$. The probability of being scattered $N$ times is $\tau_e^N$. The multiple scattering creates a power law continuum that may reach as far up in energy as the X-ray regime, the slope mainly being determined by $\tau_e$ and $T_e$ (Fransson 1982, 1984). We calculate this component by a Monte Carlo method, which takes relativistic effects, spherical geometry and the actual shock structure into account (cf. Lundqvist & Fransson 1988). On day 7, $\tau_e = 0.041 - 0.056$ for $\dot{M}_{-5}/v_{w1} = 4$, $n(\mathrm{He})/n(\mathrm{H}) = 0.1$ and $1.5 \lesssim s \lesssim 2$. In Table 4 we give the slope of the spectrum and the Compton to free-free ratio of the luminosity in the ROSAT range for $\tau_e = 0.05$, $s = 2$ and for different values of $T_e$. To estimate the free-free luminosity we use equation (3.18) with $n(\mathrm{He})/n(\mathrm{H}) = 0.1$. For low values of $T_e$ most of the Comptonized flux falls below 300 eV where the interstellar optical depth is greater than unity, assuming a galactic column density of $\sim 4.3 \times 10^{20}$ cm$^{-2}$ (Zimmermann et al. 1994). Therefore, we have listed in Table 4 the Compton to free-free ratio corrected for this column density. For other values of $s$ the Compton to free-free ratio scales as $\propto (3-s)^{-1}$. Regardless of this, it is clear that for electron temperatures higher than $\sim 1.5 \times 10^9$ K, the Comptonized flux dominates the observable flux in the ROSAT range. This is clearly incompatible with the flat spectrum observed by ROSAT (Zimmermann *et al.* 1994). In figure 7 we show the full spectrum, including Comptonization, for $T_e = T_{ion} \approx 5 \times 10^9$ K and $T_e = T_{Coul} \approx 1.5 \times 10^9$ K on day 10, clearly demonstrating the difference. *From the absence of X-rays due to Comptonization one can therefore rule out equipartition in the circumstellar shock.*

The luminosity from the reverse shock is high. For a mass loss rate of $\dot{M}_{-5}/v_{w1} = 4$ we get luminosities in the range $(1.9 - 3.6) \times 10^{41}$ erg s$^{-1}$ at 10 days, for $s = 1.5 - 2.0$ and $n \approx 25 - 30$. On day 12 the bolometric luminosity was $\sim 2 \times 10^{42}$ erg s$^{-1}$ (Lewis *et al.* 1994; Richmond *et al.* 1994).

The most characteristic aspect of the radiative scenario is the evolution of the X-ray flux. Because electron-ion equipartition may not be achieved for the circumstellar shock, and because relativistic effects are important for the free-free emission, the time evolution of the X-rays differs from that given by equation (3.13). As long as the cool shell is optically thick the spectral luminosity evolves as given by equation (3.18),

$$L(E) \propto T_e^{0.16} \, t^{(3-2s)(n-3)/(n-s)}. \qquad (3.21)$$

For $n \gg 10$, $L(E) \propto t^{-(2s-3)}$, so that for a circumstellar density gradient approaching $s = 1.5$ the luminosities in the ROSAT and ASCA bands are nearly constant with time. The observed slow decline up to $\sim 40$ days, $L(E) \propto t^{-0.30}$ (Zimmermann *et al.* 1994), is reproduced with $s \approx 1.65$.

After $\sim 100$ days the soft X-ray flux from the reverse shock begins to penetrate the cool shell. The column density of the absorbing shell decreases as $N_{cool} \propto t^{-(n-3)(s-1)/(n-s)}$, as long as the reverse shock is radiative. For $n \gg 10$, $N_{cool} \propto t^{-(s-1)}$. As $\tau(E \approx kT_{rev})$ becomes less than unity the total soft X-ray luminosity evolves according to equation (3.17). When $t \lesssim t_{cool}$, with $t_{cool}$ being given by equation (3.7), the flux decreases.

An important piece of information in favor of the cooling scenario came from the ROSAT observation from November 1993 (Zimmermann *et al.* 1993; 1994). The temperature of the X-ray emission on November 1–2 1993 was found to be only $\sim 0.5$ keV. This estimate is uncertain because line emission dominates below $\sim 2 \times 10^7$ K. Here we propose that this emission originates from the reverse shock, requiring $n \gtrsim [15.1\,(3-s)\,V_4 + s]$. This implies a very steep gradient; for $V_4 \approx 1.5$, which is reasonable at 220 days, we get $n \approx 24.6$ for $s = 2$ and $n \approx 32.2$ for $s = 1.65$. Equation (3.14) shows that the reverse shock is radiative at 220 days for all $\dot{M}_{-5}\,v_{w1}^{-1} \gtrsim 2$, $V_4 \lesssim 2$ and $n \gtrsim 15$.

In November 1993, the expected column density of the cool gas behind the shock is

$$N_{cool} = 2.6 \times 10^{20} \, \frac{(n-4)\,\dot{M}_{-5}}{v_{w1}} \, \text{cm}^{-2}, \qquad (3.22)$$

for $s = 2$, and a factor 2.7 higher for $s = 1.5$. In both cases $V_4 = 1.5$ has been assumed. Zimmermann *et al.* (1994) find $N_{cool} \approx 6 \times 10^{20}$ cm$^{-2}$. For $n \gtrsim 15$ and $\dot{M}_{-5}\,v_{w1}^{-1} \gtrsim 2$ equation (3.22) gives $N_{cool} \gtrsim 1 \times 10^{22}$ cm$^{-2}$. The $s = 1.5$ case is a factor $\sim 3$ higher. Both estimates are considerably higher than is observed.

However, Chevalier, Blondin & Emmering (1992) find that the interaction region is unstable, and it is likely that the absorption column density can be appreciably lower than estimated from equation (3.22). Typically, Chevalier *et al.* find a factor $\sim 3$ variation in the column density for different lines of sight. Also, the circumstellar density may decrease faster at large radii, compared to the $r^{-1.5}$ density law (see §2).

On day 220 the total luminosity from the reverse shock is approximately given by

$$L_{rev} \approx \begin{cases} 7.3 \times 10^{39} \, \frac{\dot{M}_{-5}}{v_{w1}} \, V_4^3 \;\; \text{erg s}^{-1} & (s = 2) \\ 5.0 \times 10^{40} \, \frac{\dot{M}_{-5}}{v_{w1}} \, V_4^{3.5} \;\; \text{erg s}^{-1} & (s = 1.5), \end{cases} \qquad (3.23)$$

where we have taken $n = 20$ in both cases. $L_{rev}$ is fairly insensitive to $n$, but sensitive to $s$. We estimate the observed luminosity by Zimmermann *et al.* to be $\sim 1 \times 10^{39}$ erg s$^{-1}$, lower than from equation (3.23) for reasonable $\dot{M}_{-5} \, v_{w1}^{-1}$ and $V$. Because of the high column density a large fraction of the luminosity is absorbed (see below), so the discrepancy is expected. Much of the remaining luminosity may be emitted as high velocity optical and UV lines from the ejecta and the cool, absorbing shell. Especially in the $s = 1.5$ case the soft X-ray luminosity may be disturbingly high compared to the bolometric luminosity.

Figure 8 shows the evolution of the calculated ROSAT, ASCA and OSSE luminosities for the $s = 2$, $n = 25$, $\dot{M}_{-5}/v_{w1} = 3.5$, $T_e = T_{Coul}$ case. While the OSSE flux decreases steadily with time, especially the ROSAT flux undergoes a dramatic change at $\sim 65$ days, shown in figure 8 as a minimum and then a slight increase of the light curve. The flattening is a result of the gradually increasing transparency of the cool shell: $N_H \propto t^{-1}$ for $n \gg 10$. The flux from the reverse shock, as shown by the dashed line, contributes an increasing fraction to the total flux in the ROSAT band.

The spectra of the $T_e = T_{Coul}$ and $T_e = T_{ion}$ models at 10 days and 200 days are shown in figure 9. While the 10 day spectrum is a smooth, free-free one-temperature spectrum from the circumstellar shock, the 200 day spectrum is completely dominated by the soft X-rays from the reverse shock. A large number of lines are seen, the strongest due to Fe XI–XIII and Ne X at $\sim 0.91 - 1.02$ keV, Mg XI–XII at $\sim 1.35 - 1.48$ keV and Si XIII at $\sim 1.87$ keV. Below $\sim 1$ keV photoelectric absorption due to the oxygen K-shell and iron L-shell gives an abrupt cutoff. The cutoff energy decreases with time according to equation (3.16). After the cool shell has become transparent the calculated ROSAT flux above the cutoff energy remains approximately constant, as long as the reverse shock is radiative and the density gradient is constant. We also note the increased extension of the high energy tail at 10 days for the $T_e = T_{ion}$ case.

As with the adiabatic models, the calculated ROSAT luminosity declines too fast for the $s = 2$ model, and a less steep density gradient is indicated. Figure 10 shows the light curves for $\rho \propto r^{-1.7}$, $n = 30$ and $T_e = T_{Coul}$, and an improved agreement with observations over the first 100 days is found. Because of the higher temperature in the $s = 1.7$ model the flux in the OSSE range has increased considerably, compared to that in the $s = 2$ model.

### 3.3. Summary of hydrodynamics

We favor the radiative case, compared to the adiabatic case, for several reasons. The density gradient of the ejecta does not need to be uncomfortably small, as is the case for the adiabatic scenario. The value of $n$ needed in the radiative case is close to that required to reproduce the radio light curves (§ 2), and the lower limit to $n$ implied by the VLBI observations (Bartel *et al.* 1994). The ROSAT/OSSE ratio is considerably easier to reproduce, and the mass loss rate derived from the X-rays is approximately the same as that derived from radio. In the adiabatic case the discrepancy is at least a factor two. To reproduce the OSSE flux the adiabatic models require substantial collisionless heating. The absence of Comptonized X-rays, however, rules out equipartition in the circumstellar shock, because this would give $T_e \gtrsim 5 \times 10^9$ K. Finally, it is extremely difficult to explain the soft X-rays at 220 days in the adiabatic model, while they are a natural consequence of the radiative model. Possible problems with the cooling scenario may be the high bolometric luminosity, and the absence of broad emission lines. However, as noted in §2, there may be a transition to a steeper density gradient in the circumstellar medium at $\sim (2-3) \times 10^{16}$ cm. This would reduce the luminosity from the reverse shock (eqn. [3.23]) after $100 - 150$ days, thus making it more likely to be compatible to the observed bolometric luminosity.

Similar calculations for the X-ray emission to ours have been done by Suzuki *et al.* (1993). Their conclusions that a high velocity and shallow gradient are necessary differ from ours. The main reason for this is that they only consider the X-ray observations during the first months. From our discussion above, the adiabatic and radiative cases do not differ substantially at these epochs, although the OSSE/ASCA ratio is more easily satisfied in the radiative case. However, a shallow density gradient is difficult to invoke in order to explain the November X-ray observation, and would probably also have problems with avoiding too high an X-ray flux due to Comptonization during the first weeks. The mass density Suzuki *et al.* use, $\dot{M}_{-5}/v_{w1} \lesssim 1$, is a factor $\gtrsim 4$ lower than we deduce from the radio observations, whereas the initial shock radius they use, $2 \times 10^{14}$ cm, is considerably larger than ours, $\sim 5 \times 10^{13}$ cm. Both a higher mass loss rate and a smaller initial radius make cooling important, and we find that the $n = 12$ model by Suzuki *et al.* is incompatible with the absence of X-ray absorption (and marginally also the $n = 8$ model). The general features of the light curve, as well as the relative ratio between the calculated ROSAT and ASCA fluxes, agree.

## 4. TEMPERATURE AND IONIZATION OF THE CIRCUMSTELLAR MEDIUM

The temperature of the circumstellar matter is crucial for estimating the mass loss rate from the radio observations, and the ionization structure must be known to calculate the absorption of the X-rays. We are also interested in the formation of emission and absorption lines in the gas. To calculate this we use the same method as was used earlier to model the circumstellar structure of SN 1979C and SN 1980K, as well as the ring structure around SN 1987A (Lundqvist & Fransson 1988, 1991).

## 4.1. *Evolution of temperature and ionization*

Ionizing radiation is emitted during breakout of the supernova shock, as well as from the gas enclosed by the reverse and circumstellar shocks. Calculations of the radiation accompanying the shock breakout for red supergiants with radii $(3-8) \times 10^{13}$ cm have been done by Falk (1978) and Klein & Chevalier (1978), who found an effective temperature immediately after the breakout of $(1-2) \times 10^5$ K, with a total energy of $\sim 3 \times 10^{48}$ ergs in a few hours. Specific calculations for SN 1993J, assuming LTE and no scattering effects, by Shigeyama et al. (1994) result in a peak temperature $\sim 2.9 \times 10^5$ K and a total energy of $\sim 10^{49}$ ergs. After $\sim 2$ days the radiation temperature has decreased to below $\sim 2 \times 10^4$ K, and the ionizing effects are small. Thereafter, the photospheric effective temperature of this model is roughly in accord with IUE observations during the first days (Fransson & Sonneborn 1994) and ground-based observations (e.g., Lewis *et al.* 1994) at later times. As soon as the shocks form the ionizing radiation is dominated by Compton upscattering of the photospheric photons by the circumstellar shock wave (see §3.2). The circumstellar structure is calculated using a time dependent photoionization code, including all interesting elements and ionization stages. Recent updates to this code are summarized in Lundqvist, Fransson & Blondin (1994). Initially, the optical depth through the circumstellar gas at 13.6 eV is $\sim 10^7$, too large to be handled by the full code. During this period, recombination can be ignored, and we follow the ionization using an adaptive mesh method similar to that described in Lundqvist & Fransson (1991).

The recombination time of hydrogen in the circumstellar gas (assuming $T = 10^5$ K) is $t_{rec} \sim 3.5 \, \dot{M}_{-5}^{-1} \, v_{w1} \, r_{15}^s$ days, where $r_{15}$ is the radius in $10^{15}$ cm. If ionized only by the radiation from the supernova breakout, the gas inside a radius of $\sim 10^{15}$ cm has time to recombine before the shock hits the gas. In this case, line emission from ions like N V, Fe X and Fe XIV is expected to be present during this period. The observed light curves of the lines are smeared out by the finite light traveling time, $t_{echo} = 2 \, r/c \approx 0.77 \, r_{15}$ days. However, the circumstellar gas is prevented from recombining by the radiation from the interaction region if the circumstellar and reverse shocks form prior to $t_{rec}$. Exactly when shock formation occurs is uncertain (Epstein 1981). If preacceleration of the gas by the radiation accompanying the shock breakout is efficient (see §4.4), a viscous shock may only develop after about one expansion time. Otherwise, it may form earlier. The optical depth to electron scattering of the circumstellar gas from radius $R$ out to infinity for $s = 2$ is unity for $R_{\tau_e=1} \approx 2 \times 10^{13} \dot{M}_{-5} \, v_{w1}^{-1}$ cm. The scattering photosphere may therefore be located at a considerable distance out in the circumstellar gas. This delays both the escape of the radiation and the formation of the viscous shock. In the case of $s = 1.5$, or a cavity around the progenitor formed by the influence of a binary, the delay of the burst is not important.

We have calculated the ionization and temperature structure of the circumstellar gas for two models: one has $\dot{M}_{-5}/v_{w1} = 5$, $s = 1.7$ and $n = 30$, and the other $\dot{M}_{-5}/v_{w1} = 4$, $s = 2$ and $n = 25$. Both give good fits to the X-ray emission as discussed in §3.2. We adopt $n(\text{He})/n(\text{H}) = 0.3$, $n(\text{N})/n(\text{O}) = 1.5$ and $n(\text{N})/n(\text{C}) = 3.5$ as was found for the circumstellar gas around SN 1987A (Lundqvist *et al.* 1994). Other elements have solar abundances, and the overall metallicity is also solar. For the first two days we take the evolution of the photosphere from Shigeyama *et al.* (1994). Because the color temperature is likely to be larger than the effective temperature, we have multiplied the latter by a correction factor $q = T_{color}/T_{eff}$, retaining the bolometric luminosity in the model by Shigeyama *et al.* . Klein & Chevalier (1978) found in their models that $q \lesssim 2$, and Ensman & Burrows (1992) found $q \sim (1.0 - 1.8)$ for SN 1987A. Here we use $q = 1.35$. We also include the time delay between the ionizing photons from the limb of the supernova and its center. For times later than two days we have joined the model by Shigeyama *et al.* to IUE (Fransson & Sonneborn 1994) and ground-based observations (Lewis *et al.* 1994). The radiation from the circumstellar and reverse shocks is included as discussed above, assuming the electrons of the gas behind the circumstellar shock are heated by Coulomb collisions only. Furthermore, we assume that the viscous shock forms within the first expansion time scale, i.e., prior to $\sim 0.23$ days. The exact time of shock formation is not important for the temperature and ionization structures of the circumstellar gas at observable times as long as the viscous shock does not form later than after $\sim 0.5$ days (see below).

During the supernova breakout, hydrogen and helium in the circumstellar gas quickly become fully ionized; even at $10^{16}$ cm the ionization takes less than 100 s. Carbon, nitrogen and oxygen all attain their He-like stages, and iron close to the supernova is ionized up to Fe XVII. The temperature of the circumstellar gas, $T_{wind}$, becomes fairly constant, $\sim 3 \times 10^5$ K, throughout the wind, scaling roughly linearly with the adopted value of $q$. $T_{wind}$ is also a weak function of the He/H-ratio; for $n(\text{He})/n(\text{H}) = 0.1$ instead of 0.3, $T_{wind}$ is reduced to $\sim 2.8 \times 10^5$ K. As the color temperature of the burst decreases, the gas close to the supernova cools by Compton cooling. Further out in the circumstellar gas the initial structure is more or less frozen-in. If the viscous shock forms after $\sim 0.5$ days, the photospheric flux above 100 eV is low enough for recombination to N V close to the ejecta to become important. In our models, shock formation occurs earlier and recombination is negligible; the circumstellar gas becomes completely ionized by the radiation from the shocks, once these have formed. At the same time, $T_{wind}$ increases rapidly.

In figure 11 we show the evolution of $T_{wind}$ at $1.5 \times R_s$, which is roughly the temperature to be used in equation (2.3). The peak is due to Compton heating of the circumstellar gas by the radiation from the circumstellar shock. This radiation is during the first days totally dominated by Comptonization of photospheric photons, and completely ionizes the circumstellar gas. The flux due to Comptonization decreases rapidly with time, and as a result so does also the temperature of the circumstellar gas close to the shock. After $\sim 10 - 15$ days, the temperature remains roughly constant at a level slightly in excess of $10^5$ K. The ionization structure evolves less dramatically; around day 5, the gas is still essentially fully ionized, except for iron which is dominated by Fe XIV. After $\sim 5$ days, high populations of hydrogen-like C, N and O start to build up, especially in the $s = 1.7$ model. This has important effects on the circumstellar X-ray absorption (see §4.2). After $\sim 30$ days the degree of ionization stays roughly constant. At 200 days, the $s = 1.7$ model is dominated by roughly equal populations of fully ionized and hydrogen-like stages of carbon and nitrogen, whereas oxygen is mainly in O VII–VIII and iron in Fe VIII–IX. The gas in the $s = 2$ model has recombined even less. An important difference between the two models is that the zone with partial recombination is more extended in the $s = 1.7$ model, reaching out to $\sim 15 \times R_s$, whereas in the $s = 2$ model it is mainly gas inside $5 \times R_s$ that is affected by recombination.

### 4.2. Circumstellar X-ray absorption

The column density of the circumstellar gas is for $n(He)/n(H) = 0.1$ and $\xi = 1.3$ (§2) given by

$$N(H)_{cs} = 2.1 \times 10^{22} \; \frac{\dot{M}_{-5}}{(s-1) \, v_{w1}} \; V_4^{1-s} \left( \frac{t}{8.90 \text{ days}} \right)^{1-s} \text{ cm}^{-2}. \qquad (4.1)$$

ROSAT observations show the total column density to be consistent with the galactic absorption in the M 81 direction, $\sim 4.3 \times 10^{20}$ cm$^{-2}$ (Zimmermann *et al.* 1994). With $t \approx 10$ days and $V_4 \approx 2$ we find $\dot{M}_{-5}/v_{w1} \lesssim 0.045$ (0.025) for $s = 2.0$ (1.7). These values are clearly incompatible with the value of $\dot{M}/v_w$ derived from the radio and X-ray observations, unless the circumstellar gas is almost completely ionized so that K-shell electrons of C, N and O, and L-shell electrons of Fe are removed. In figure 12 we show the optical depth through the circumstellar gas at 1 keV, $\tau(1 \text{ keV})$, as a function of time for the $s = 1.7$ and $s = 2.0$ models. It is seen that with the fading of the radiation due to Comptonization, $\tau(1 \text{ keV})$ increases rapidly and peaks around day $10 - 20$ at $\tau(1 \text{ keV}) \sim 0.3$ (1.5) for $s = 2.0$ (1.7). This is higher than $\tau(1 \text{ keV}) \sim 0.1$ inferred from $N(H) \sim 4.3 \times 10^{20}$ cm$^{-2}$. The high value we find for the $s = 1.7$ model can be suppressed if the overall metallicity is lower than solar, the $n(N)/n(O)$ value is higher than 1.5, or if the $\rho_w \propto r^{-1.7}$ dependence steepens at large radii, as may be indicated by observations at 3 mm (Phillips & Kulkarni 1993b). $\tau(1 \text{ keV})$ may also decrease if the circumstellar gas is non-spherically symmetric.

As was discussed in §4.1, the temperature and ionization structures stay roughly constant after $\sim 20 - 30$ days. As a result, the X-ray optical depth through the circumstellar gas decreases with time after $\sim 20$ days (figure 12). We therefore think it is unlikely that the observed increase in column density of absorbing gas up to $\sim 200$ days is due to recombination of the circumstellar gas, as has been advocated (Zimmermann *et al.* 1994). As discussed in §3.2, we believe the softening of the X-ray spectrum is instead a result of the cool shell between the circumstellar and reverse shocks becoming transparent.

### 4.3. Implications for the radio absorption and the optical and UV line emission

Our calculations give very high circumstellar temperatures close to the shock for $t \lesssim 10$ days (fig. 11). This reduces $\tau_{ff}$ (eqn. [2.3]) so that $\tau_{ff}$ becomes less than unity at 1.3 and 2 cm at all times, in contradiction to the observations. Our calculations also show too low a population of N V to be compatible with the IUE observations. These problems may be solved in either of two ways: a low-density region close to the supernova (cavity), decreasing the early circumstellar emission from the shocks, or alternatively, clumping of the circumstellar medium.

Clumping will lead to a higher N V $\lambda$ 1240 luminosity than in our model, though only if the ionizing luminosity does not increase appreciably with increasing circumstellar density. If the clumps have a filling factor, $f_{cl}$, and the ionizing luminosity is kept at a constant level, the line emission increases roughly as $\propto f_{cl}^{-1}$ compared to that for a smooth wind, while the effect of clumping on the ionizing luminosity is more uncertain. If the clumps have a spherically symmetric distribution, and they dissolve quickly behind the circumstellar shock, the ionizing flux should be similar to that for a smooth wind. If the clumps do not dissolve efficiently, the free-free to Compton ratio should be higher. The maximum ratio could be $\gtrsim f_{cl}^{-1}$, and occurs for very large clumps, or a disk. From numerical calculations we find that when the contribution due to Comptonization is kept at the same level as in §4.1 a high enough N V $\lambda$ 1240 luminosity is produced if $f_{cl}$ is of the order $10^{-2}$.

Clumping also affects the circumstellar absorption. In particular, $\tau_{ff} \propto \rho^2$, which means that as long as the wind is optically thick, $\tau_{ff} \propto f_{cl}^{-1}$. If the ionizing flux is kept fixed, $\tau_{ff}$ increases even faster with decreasing $f_{cl}$, since a higher density results in faster cooling of the circumstellar gas, and hence an increase of the free-free absorption (eqn. [2.3]). The clumpy scenario may thus give rise to both N V zones in the circumstellar gas, and also explain the absence of detected radio emission during the first $\sim 5$ days. If $f_{cl}$ is independent of radius, we still need an overall distribution $r^{-s}$, with $s < 2$, to explain the shape of the radio and X-ray light curves.

A clue to whether clumping was important for SN 1993J may come from the observed H$\alpha$ flux. Using the results by Cumming $et$ $al.$ (1994), and $E_{B-V} = 0.18$ (Fransson & Sonneborn 1994), one finds that the dereddened flux on day 2.2 corresponds to an H$\alpha$ luminosity of $L_{H\alpha} \approx 1.2 \times 10^{38}$ erg s$^{-1}$. For a spherically symmetric wind, this means that if the gas emitting the line resides in a region between $R_s$ and $r$, the mass loss needed to explain the H$\alpha$ emission corresponds to

$$\frac{\dot{M}_{-5}}{v_{w1}} \approx 33.1 \, f_{cl}^{0.5} \, (2s-3)^{0.5} \, T_5^{0.47} \left(\frac{R_s}{10^{15} \text{ cm}}\right)^{s-1.5} \left(1 - \left(\frac{r}{R_s}\right)^{3-2s}\right)^{-0.5}, \quad (4.2)$$

assuming Case B emissivity (e.g., Osterbrock 1989), $n(\text{He})/n(\text{H}) = 0.3$ and $s > 1.5$. The maximum radius, $r$, is approximately given by the light echo radius, $r = ct/2$. For $T = 10^5$ K, $f_{cl} = 1$ and $t = 2.2$ days, equation (4.2) gives $\dot{M}_{-5}/v_{w1} \approx 22 - 24$ for $1.6 \lesssim s \lesssim 2$. This is a factor of $\sim 10$ higher than used in §4.1. A similar value is found for N V $\lambda$ 1240, if the ionizing luminosity is kept at the same level as in §4.1.

Comptonization is also suppressed if the progenitor was surrounded by a low-density region, since $\tau_e \propto M_{cs} = \dot{M}(R_s - R_o)/v_w$ ($s = 2$), where $M_{cs}$ is the swept up circumstellar mass and $R_o$ is the radius of the low-density region. Slower Coulomb heating leads to a lower electron temperature. If the low-density region extends to $\sim (0.5 - 1) \times 10^{15}$ cm, the early high level of Comptonization is avoided, giving a lower circumstellar temperature outside this radius than in §4.1, and hence a higher radio absorption. In the case of a low-density region, the radio flux is lower also because the radio emissivity scales as $\propto \rho_w^{(3+\gamma)/2}$. A low-density region therefore can explain the observed onset of the radio emission after $\sim 5$ days. Unless the wind exterior to the low-density region is clumpy, it is necessary in this scenario to invoke radiative excitation to explain the observed strong N V $\lambda$ 1240 and H$\alpha$ flux. While the N V line should be dominated by direct radiative excitation, H$\alpha$ is probably boosted also by excitation to the n = 3 level by L$\beta$. We find that a very steep spectrum due to Comptonization is necessary for radiative excitation to operate; at 2.2 days, $\alpha \sim 4 - 5$. This implies that if radiative excitation is important, a very low density close to the progenitor is necessary; the low-density region would essentially resemble a cavity.

### 4.4. *Preacceleration*

The strong burst of radiation during the first few days may have important dynamical effects on the circumstellar gas. Assuming that electron scattering dominates the opacity, an initial UV-burst of $10^{49} E_{49}$ erg s$^{-1}$ preaccelerates the circumstellar gas to

$$v_{cs} = 1.42 \times 10^4 \ E_{49} \ V_4^{-2} \left(\frac{R_s}{r}\right)^2 t_{days}^{-2} \ \text{km s}^{-1} \qquad (4.3)$$

where $R_s$ is the shock radius and $r$ the distance from the supernova. With $V_4 \approx 2$ the preshock velocity on day 2 is therefore $\sim 900 \ E_{49}$ km s$^{-1}$, decreasing with time as $t^{-2}$. Observations of H$\alpha$ emission on March 29, 1993 show evidence for velocities of $\sim 1000$ km s$^{-1}$ (Cumming *et al.* 1994). Also the N V $\lambda \lambda$ $1238 - 1242$ lines indicate similar velocities (Fransson & Sonneborn 1994). We have integrated the energy from the light curve model for SN 1993J by Shigeyama *et al.* (1994), and find up to day 2.0, $E = 6 \times 10^{48}$ ergs. In addition to the radiation from the photosphere, the circumstellar gas is also accelerated by the radiation from the circumstellar shock. In the $s = 2.0$ (1.7) model this contribution to $E$ amounts to $\sim 5 \times 10^{48}$ ($\sim 2 \times 10^{48}$) ergs up to day 2.0. Together, the flux from the photosphere and that from the circumstellar shock may be sufficient to explain the observed preacceleration. In the adiabatic model, $n$ is small and $V \propto t^{-(3-s)/(n-s)}$ is large at very early times (eqn. [2.2]). Therefore, it is more difficult to obtain enough preacceleration in the adiabatic model, which provides an additional, indirect piece of evidence in favor of the radiative models.

Line acceleration may also be important (Fransson 1986). Although the efficiency requires a detailed study, it was argued that the lines can increase the acceleration by a factor $4 - 10$, compared to electron scattering. Line acceleration requires that the wind is not completely ionized, which may be fullfilled initially inside $\sim 10^{15}$ cm, and certainly has to be the case in the gas emitting the N V lines. Because preacceleration is strongly dependent on radius, the line widths are expected to decrease with time. Fransson (1986) argued that the velocity should depend on radius as $r^{-3.3}$, and the line widths therefore decrease roughly as $t^{-3.3}$. If the circumstellar gas is accelerated over an extended period of time, the line widths may decrease somewhat more slowly.

## 5. DISCUSSION

The $r^{-1.5}$ model shown in figures 1 and 2, and the models for the X-ray emission discussed in §3, indicate that there is evidence for a shallower density dependence in the circumstellar gas than an $r^{-2}$ law. The drop in the luminosity at 3 mm after $\sim 100 - 150$ days (Phillips & Kulkarni 1993b; §2) indicates that the departure from an $r^{-2}$ dependence only refers to the region traversed by the shock between $\sim 8$ and $\sim 150$ days. For an $r^{-1.5}$ dependence this translates into a variation in the mass loss rate by a factor of $\sim 4$ over this region. Variations in the density on even shorter length scales are indicated by bumps in the optically thin part of the 2 cm light curve after $\sim 50$ and $\sim 100$ days (Pooley & Green 1993; Weiler $et\ al.$ 1994), and of the 3.6 cm light curve after $\sim 100$ days (Weiler $et\ al.$ 1994).

A density dependence different from $\rho_w \propto r^{-2}$ could be the result of either a variation of the mass loss rate on a time scale $\sim R_s/v_w = V_s\ t/v_w \approx 700\ v_{w1}^{-1}$ years before the explosion (equivalent to a length scale of $\sim 2 \times 10^{16}$ cm, see §2) where $R_s$ is the shock radius, or a similar variation in the wind velocity. A possible origin of this may be connected to instabilities in the nuclear burning during or after carbon burning (e.g., Woosley, Weaver & Taam 1980). The conditions for this to occur are, however, restricted to He-cores between $2.5 - 2.8\ M_\odot$. Another possibility is that the mass loss is due to pulsational instabilities in the envelope. However, the time scale for the fundamental pulsational mode in red supergiants is typically $\sim 10$ years (Fox & Wood 1982), corresponding to radio variations on a time scale of $\sim 1.5$ days. Because variations on time scales less than $R_s/c = Vt/c \approx t/15$ are smoothed out by light travel delays, the shortest mass loss variations that can be detected at $\sim 50$ days therefore correspond to $\sim 20$ years. Alpha Orionis shows evidence for periodicity in its mass loss on a time scale of 5.8 years and for structure in its circumstellar envelope on a scale of $\sim 10^{16}$ cm (Goldberg 1979).

If the progenitor was a member of a binary system, this may modify the circumstellar environment considerably. Woosley $et\ al.$ (1994) estimate a binary separation $a \approx 1.5 \times 10^{14}$ cm for SN 1993J. With a total mass of $\sim 20\ M_\odot$ this corresponds to a period of about 7.1 years, and an orbital velocity of $\sim 45$ km s$^{-1}$. If the companion star also possesses a wind this may further complicate the density structure. An early type companion, as may be indicated by the observations by Aldering $et\ al.$ (1994), would have a fast wind with velocity $\sim 1000$ km s$^{-1}$ and mass loss $10^{-7} - 10^{-6} M_\odot$ yr$^{-1}$. For two early-type components Stevens, Blondin & Pollock (1992) find that a dense interface forms due to radiative cooling of the gas shocked by the colliding winds. The dense, shocked gas may be helpful in explaining the N V $\lambda$ 1240 and H$\alpha$ emission discussed in §4.3. A structure similar to that in the calculation by Stevens $et\ al.$ is also seen in a model for the interacting winds in symbiotics (Nussbaumer & Walder 1993), which invoke one hot star and one red giant, and may even better represent the situation for SN 1993J. The momentum of the fast wind from the companion may dominate the slow wind from the supernova progenitor. For $\dot{M}_w \sim 4 \times 10^{-5} M_\odot$ yr$^{-1}$ and $v_w \sim 10$ km s$^{-1}$, this occurs for a mass loss rate of the companion, $\dot{M} \gtrsim 4 \times 10^{-7} M_\odot$ yr$^{-1}$, assuming a wind speed of $\sim 1000$ km s$^{-1}$. It is likely that the wind structure was modified also by the orbital motion of the system; the wind density should be higher in the orbital plane than orthogonal to this, producing a disk-like structure with thickness depending on details of the wind interaction and the orbital motions of the two stars. To explore fully the interaction of the ejecta with a disk-like structure, multidimensional hydrodynamics is necessary. However, assuming a freely expanding circumstellar shock, Lundqvist (1994) argues that a disk structure may explain the radio light curves, without invoking a time varying mass loss from the progenitor, if the disk thickness increases roughly as $\propto r^{0.4}$. To model accurately the right turn-ons at different wavelengths, and since this requires a substantial line-of-sight through the circumstellar gas exterior to the circumstellar shock, it is unlikely that the tentative disk was thin.

In §4.3, we found evidence for that the wind close to the progenitor may have been clumpy, or that it was surrounded by a low-density region. Because clumping, a cavity, and $\rho_w \propto r^{-s}$ ($s < 2$) may be natural consequences in the binary scenario, we find it plausible that the progenitor of SN 1993J was a member of a binary. Asymmetry may also decrease somewhat the path length of the line-of-sight through the circumstellar gas, thus explaining the low absorbing column density inferred from the ROSAT observations. However, VLBI observations during the first $\sim 50$ days (Bartel $et\ al.$ 1994) showed no sign of asymmetry, so we cannot rule out an explanation based on a time varying mass loss. Future VLBI observations may reveal which scenario is the most likely one. If $f_{cl} \ll 1$ and/or the progenitor was surrounded by an extended cavity, our conclusion in §3.2 that the absence of Comptonization during the observed epochs is evidence for $T_e = T_{Coul}$ may not hold true.

## 6. CONCLUSIONS

Here we summarize our main conclusions:

1. From the radio observations there are strong indications of a circumstellar medium with density flatter than $\rho_w \propto r^{-2}$, at least out to $\sim 2 \times 10^{16}$ cm. The observations are compatible with a radial outflow with $\rho_w \propto r^{-1.5}$, indicating either a decreasing $\dot{M}/v$ leading up to the explosion, or non-spherically symmetric mass loss. The mass loss rate from radio observations is in the range $(2-6)\times 10^{-5} M_\odot$ yr$^{-1}$ at $10^{15}$ cm for $v_w = 10$ km s$^{-1}$. The main uncertainty is the temperature of the circumstellar gas.

2. Most of the early X-ray flux is likely to originate in the shocked circumstellar gas. The reverse shock is radiative in our most likely scenario, and the flux from the postshock gas is absorbed by the cooling gas between the shock and the contact surface. The ROSAT and ASCA luminosities are compatible with the mass loss rate derived from the radio observations. The X-ray light curve indicates a flat circumstellar density, $\rho_w \propto r^{-1.7}$, similar to that derived from the radio observations.

The soft X-rays observed at 220 days are explained as emission from the reverse shock leaking out as the shell becomes transparent.

3. The level of the OSSE observations indicates an electron temperature of $\sim 1.5 \times 10^9$ K. The absence of X-rays due to Comptonization gives an upper limit to the electron temperature of $\lesssim 1.5 \times 10^9$ K for the shocked circumstellar gas. This is a factor of $\sim 4$ less than the ion temperature, and indicates that collisionless heating in the shock front is not important.

4. There are several indications of a steep density gradient in the ejecta. The turn-ons of the radio light curves, the radiative scenario for the X-rays, and the late soft X-rays all require $n \gtrsim 20$. Additional, though weaker, evidence comes from preacceleration. A steep density gradient, $n \gtrsim 20$, is consistent with both VLBI observations (Bartel *et al.* 1994) and with hydrodynamic models for the explosion (Suzuki *et al.* 1993).

5. The temperature of the circumstellar medium is $\sim (1 - 3) \times 10^5$ K for most of the time during the first 100 days. After the formation of the shocks the circumstellar medium was nearly completely ionized, in agreement with the observed lack of substantial X-ray absorption. To explain the long duration and absolute flux of the N V luminosity, as well as the flux of the H$\alpha$ line, a non-uniform density distribution may be required.

6. The high velocities indicated by the N V and H$\alpha$ lines are explained as a result of preacceleration by the radiation in connection with the break out of the shock. Radiation from the circumstellar shock also contributes to the acceleration of the circumstellar gas.

SN 1993J in many respects conforms to the same class of Type II supernovae as SN 1979C and SN 1980K. There are, however, indications from both radio, X-rays and UV that the circumstellar medium is different from a standard $\rho_w \propto r^{-2}$ stellar wind. Either a variable mass loss rate, or a non-radial outflow, may explain the density variation. The latter alternative may be expected from mass loss in a binary system, and is supported by the clumping and/or a cavity we infer from the N V $\lambda$ 1240 and H$\alpha$ lines.

Finally, there are several predictions for the future. Because of the expansion the column density of the cool gas absorbing the reverse X-rays decreases with time as in equation (3.15), and the soft X-ray cutoff is expected to decrease according to equation (3.16). The temperature of the reverse shock is $kT_{rev} \approx 1.16 \ (V/20000 \ \text{km s}^{-1})^2 \ [(n-2)/20]^{-2}$ keV. The soft X-ray emission is therefore a valuable diagnostic of the density gradient and the hydrodynamic structure of the outer ejecta. As the column density decreases the observed X-ray flux from the reverse shock is expected to increase considerably, as long as the shock is radiative and the circumstellar density does not drop more rapidly than $r^{-2}$. In the adiabatic phase it drops according to equation (3.13).

As the radioactive input from the supernova decays, and if the circumstellar density does not decrease very fast, the ingoing X-ray emission is expected to dominate the energy input to the supernova, as is the case for such supernovae as SN 1979C, SN 1980K, SN 1988Z and SN 1986J (CF94). This results in UV and optical emission lines mainly from the outer parts of the ejecta. The line widths are therefore expected to be considerably larger, $(10 - 15) \times 10^4$ km s$^{-1}$, than at earlier epochs. If the circumstellar medium is highly clumpy also low velocity lines from shocked circumstellar gas may appear. In contrast to the earlier phases, high ionization lines like C IV $\lambda$ 1550 and [O III] $\lambda$ $\lambda$ $4959 - 5007$ are strong, as well as Ly$\alpha$, H$\alpha$ and Mg II $\lambda$ 2800. This is discussed in detail in CF94. The future evolution of SN 1993J is therefore of obvious interest for understanding both the ejecta and the circumstellar gas.

Acknowledgments: We are grateful to Bernd Aschenbach, Robert Cumming, Peter Meikle, Mark Leising and Ken Nomoto for discussions, to Toshikazu Shigeyama for light curve models, and to David Green for permission to use their observations prior to publication. This research was supported by NSF Grants AST-9016687 and AST-9314724, the Swedish Natural Sciences Research Council and the Göran Gustafsson Foundation for Research in Natural Sciences and Medicine.

# FIGURE CAPTIONS

Fig. 1.— Radio light curves at 2 cm assuming a spectral index of $\gamma = 0.7$ together with observations from Pooley & Green (1993). The full line gives the "best fit" for the $\rho_w \propto r^{-2}$, $n = 30$ model, and the dashed for the $\rho_w \propto r^{-1.5}$, $n = 30$ model.

Fig. 2.— Radio light curves at 3 mm (dashed-dotted line), 1.3 cm (short-dashed line), 6 cm (solid line), and 20 cm (long-dashed line) for the $\rho_w \propto r^{-1.5}$, $\gamma = 0.7$, $n = 30$ model. Observations at 3 mm are from Phillips & Kulkarni (1993a) and those at other wavelengths from Weiler $et\ al.$ (1994)

Fig. 3.— Density, electron temperature and ion temperature for the $s = 2$, $n = 6$, $\dot{M}_{-5}/v_{w1} = 3$ model at 10 days. Only Coulomb heating of the electrons is assumed, leading to $T_e \ll T_{ion}$ for the circumstellar shock.

Fig. 4.— Spectrum at 10 days for $s = 2$, $n = 6$ and $\dot{M}_{-5}/v_{w1} = 3$. The upper panel assumes $T_e = T_{Coul}$ and the lower $T_e = T_{ion}$. The dotted lines give the contribution from the reverse shock, and the dashed lines that from the circumstellar shock.

Fig. 5.— Density, electron temperature and ion temperature for the $s = 2$, $n = 25$, $\dot{M}_{-5}/v_{w1} = 4.0$ model at 10 days. Only Coulomb heating of the electrons is assumed. The density spike close to the reverse shock is due to the cool shell discussed in the text.

Fig. 6.— The density (dashed line) and temperature (solid line) of the radiative, reverse shock for the same model as in figure 5.

Fig. 7.— Spectrum at 10 days for the $s = 2$, $n = 25$, $\dot{M}_{-5}/v_{w1} = 4.0$ model including Comptonization of photospheric photons by the nearly relativistic electrons behind the circumstellar shock. Left panel is for $T_e = T_{Coul}$ and right for $T_e = T_{ion}$. Short-dashed lines show the free-free luminosity, dotted lines the Comptonized luminosity and solid lines the total luminosity. Long-dashed lines show the observable luminosity assuming an interstellar column density of $4.3 \times 10^{20}$ cm$^{-2}$. With $T_e = T_{ion}$ especially the ROSAT range is dominated by the Compton contribution, with a power law slope of $\sim 1.0$. This is in contradiction to the observed spectrum.

Fig. 8.— Luminosity as function of time in the ROSAT $(0.1 - 2.4$ keV), ASCA $(1 - 10$ keV), and OSSE $(50–150$ keV) bands for the $s = 2$, $n = 25$, $\dot{M}_{-5}/v_{w1} = 4.0$, $T_e = T_{Coul}$ model. The dashed line gives the total luminosity from the reverse shock, while the dotted line gives the same from the circumstellar shock. The corresponding contributions to the ROSAT luminosity are also shown.

Fig. 9.— Spectrum at 10 days and 200 days for the radiative reverse shock model with $s = 2$, $n = 25$, $\dot{M}_{-5}/v_{w1} = 4.0$. The dotted lines give the contribution from the reverse shock. At 10 days the hard emission from the circumstellar shock dominates at all energies, while at 200 days the soft component from the reverse shock dominates. Most of the emission from the reverse shock is due to line emission. Below $\sim 1$ keV photoelectric absorption from the cooling gas cuts off the reverse emission. Interstellar and circumstellar absorption are not included.

Fig. 10.— Same as figure 8 for $s = 1.7$, $n = 30$, $\dot{M}_{-5}/v_{w1} = 4.5$ and $T_e = T_{Coul}$.

Fig. 11.— Temperature of the circumstellar gas at $R = 1.5 \times R_s$ for $s = 1.7$ and $\dot{M}_{-5}/v_{w1} = 4.0$ (dashed line), and $s = 2.0$ and $\dot{M}_{-5}/v_{w1} = 5.0$ (solid line).

Fig. 12.— Optical depth through the circumstellar medium at 1 keV for the two models in figure 11. Solid and dashed lines represent the same models as in figure 11.

**TABLE 1**

Cooling times for the reverse shock
for $s = 1.5$ and $s = 2$

| $n$ | $C_{n,1.5}$ days | $t_{cool}^a$ days | $C_{n,2}$ days | $t_{cool}^a$ days |
|---|---|---|---|---|
| 6 | 0.366 | 30.39 | 0.498 | 11.20 |
| 7 | 0.150 | 7.63 | 0.199 | 3.17 |
| 8 | 0.0760 | 2.76 | 0.0996 | 1.26 |
| 9 | 0.0439 | 1.25 | 0.0569 | 0.61 |
| 10 | 0.0277 | 0.65 | 0.0356 | 0.34 |
| 12 | 0.0131 | 0.23 | 0.0166 | 0.13 |

[a] $t_{cool}$ is the cooling time at 1 day for $\dot{M}_{-5}/v_{w1} = 2$ and $V_4(10 \text{ days}) = 2$.

**TABLE 2**

Luminosities at 10 days in the ROSAT, ASCA and OSSE bands for $s = 2$.

| n | $\dot{M}_{-5}/v_{w1}$ | $\rho_0$ $10^{-16}$ g cm$^{-3}$ | $T_e$ | $V_4$ | $T_{e\ rev}$ $10^8$ K | $T_{e\ cs}$ $10^8$ K | $L_{0.1-2.4}^a$ | $L_{1-10}^a$ | $L_{100}^b$ |
|---|---|---|---|---|---|---|---|---|---|
| 5 | 4.0 | 0.021 | Coul | 2.08 | 4.5 | 13.7 | 2.89 | 5.69 | 0.61 |
| 6 | 2.0 | 0.1 | Coul | 2.07 | 3.1 | 10.7 | 2.06 | 3.62 | 0.16 |
|   |   |   | eq | 2.02 | 3.1 | 45.7 | 3.07 | 5.07 | 0.47 |
| 7 | 1.0 | 0.29 | Coul | 2.00 | 2.0 | 8.7 | 1.77 | 2.25 | 0.03 |
|   | 1.5 | 0.44 | Coul | 2.01 | 2.2 | 10.5 | 5.27 | 5.18 | 0.09 |
|   |   |   | eq | 1.95 | 1.7 | 51.2 | 7.82 | 6.70 | 0.28 |
|   | 2.0 | 0.63 | Coul | 2.01 | 2.2 | 10.5 | 6.50 | 8.90 | 0.17 |

[a] in $10^{39}$ erg s$^{-1}$

[b] in $10^{38}$ erg s$^{-1}$ keV$^{-1}$

**TABLE 3**

Luminosities at 10 days in the ROSAT, ASCA and OSSE bands
for a radiative reverse shock.

| $s$ | n | $(\dot{M}_{-5}/v_{w1})^a$ | $T_e$ | $V_4$ | $T_{e\ rev}$ $10^8$ K | $T_{e\ cs}$ $10^8$ K | $L_{0.1-2.4}^b$ | $L_{1-10}^b$ | $L_{100}^c$ |
|-----|-----|------|------|------|------|------|------|------|------|
| 1.5 | 35 | 4.0 | Coul | 2.00 | 0.11 | 18.2 | 1.80 | 5.58 | 1.89 |
|     |    | 5.0 | Coul | 2.00 | 0.11 | 19.8 | 2.82 | 8.72 | 3.18 |
| 1.7 | 30 | 4.0 | Coul | 2.00 | 0.12 | 17.7 | 1.75 | 5.43 | 1.75 |
|     |    | 5.0 | Coul | 2.00 | 0.11 | 19.0 | 2.68 | 8.28 | 2.82 |
|     |    |     | eq   | 2.00 | 0.11 | 53.7 | 3.00 | 9.66 | 5.22 |
| 2.0 | 25 | 4.0 | Coul | 2.00 | 0.10 | 16.6 | 1.98 | 5.97 | 1.61 |
|     |    |     | eq   | 2.00 | 0.10 | 58.1 | 2.18 | 6.72 | 2.82 |
|     |    | 5.0 | Coul | 2.00 | 0.10 | 18.4 | 3.33 | 10.0 | 2.82 |

$^a$ at $10^{15}$ cm

$^b$ in $10^{39}$ erg s$^{-1}$

$^c$ in $10^{38}$ erg s$^{-1}$ keV$^{-1}$

**TABLE 4**

Effects of Comptonization on the ROSAT emission
at 7 days for $\tau_e = 0.05$ and $s = 2$.

| $T_{e\ cs}$ $10^9$ K | $\alpha$ | $L_{Comp}/L_{ff}$ $N_H = 0$ | $L_{Comp}/L_{ff}$ $N_H = 4.3 \times 10^{20}$ cm$^{-2}$ |
|------|------|------|------|
| 1.0 | 2.50 | 1.36 | 0.26 |
| 1.5 | 2.00 | 8.25 | 2.43 |
| 2.0 | 1.66 | 25.6 | 9.20 |
| 3.0 | 1.30 | 101. | 49.5 |
| 5.0 | 0.96 | 341. | 203. |

Fig. 1

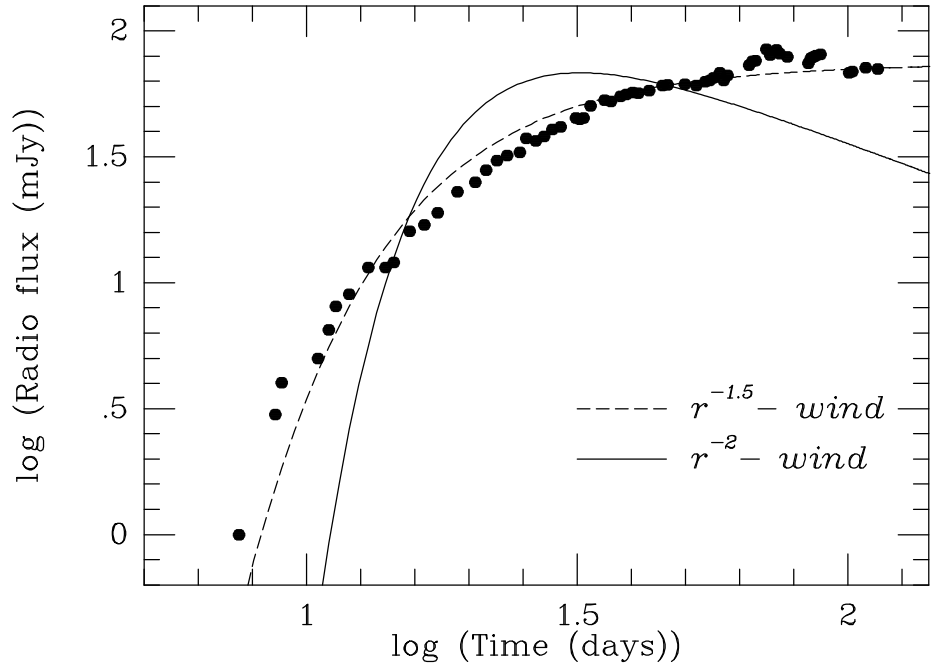

Fig. 2

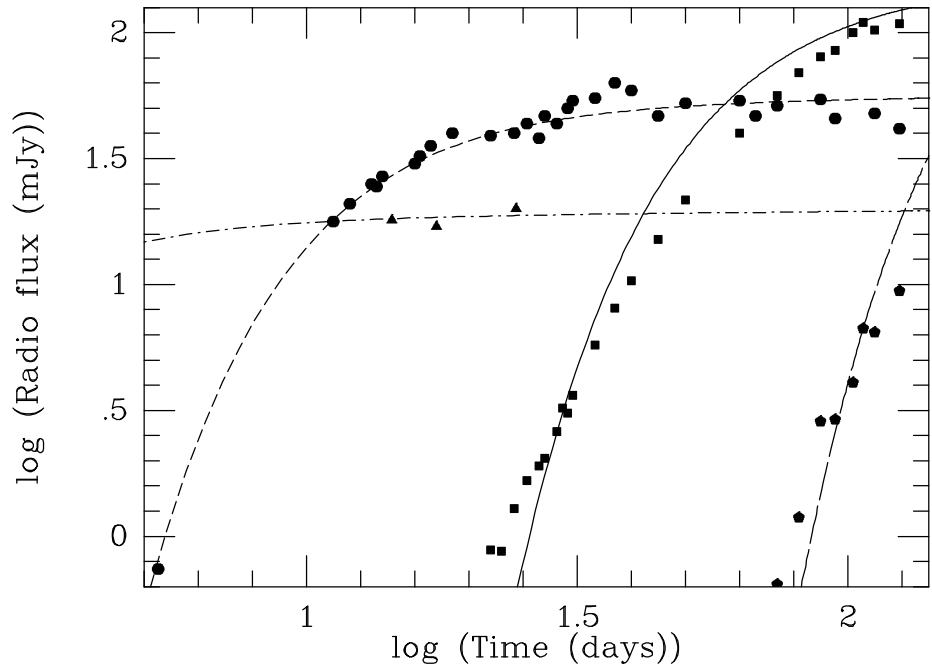

Fig. 3

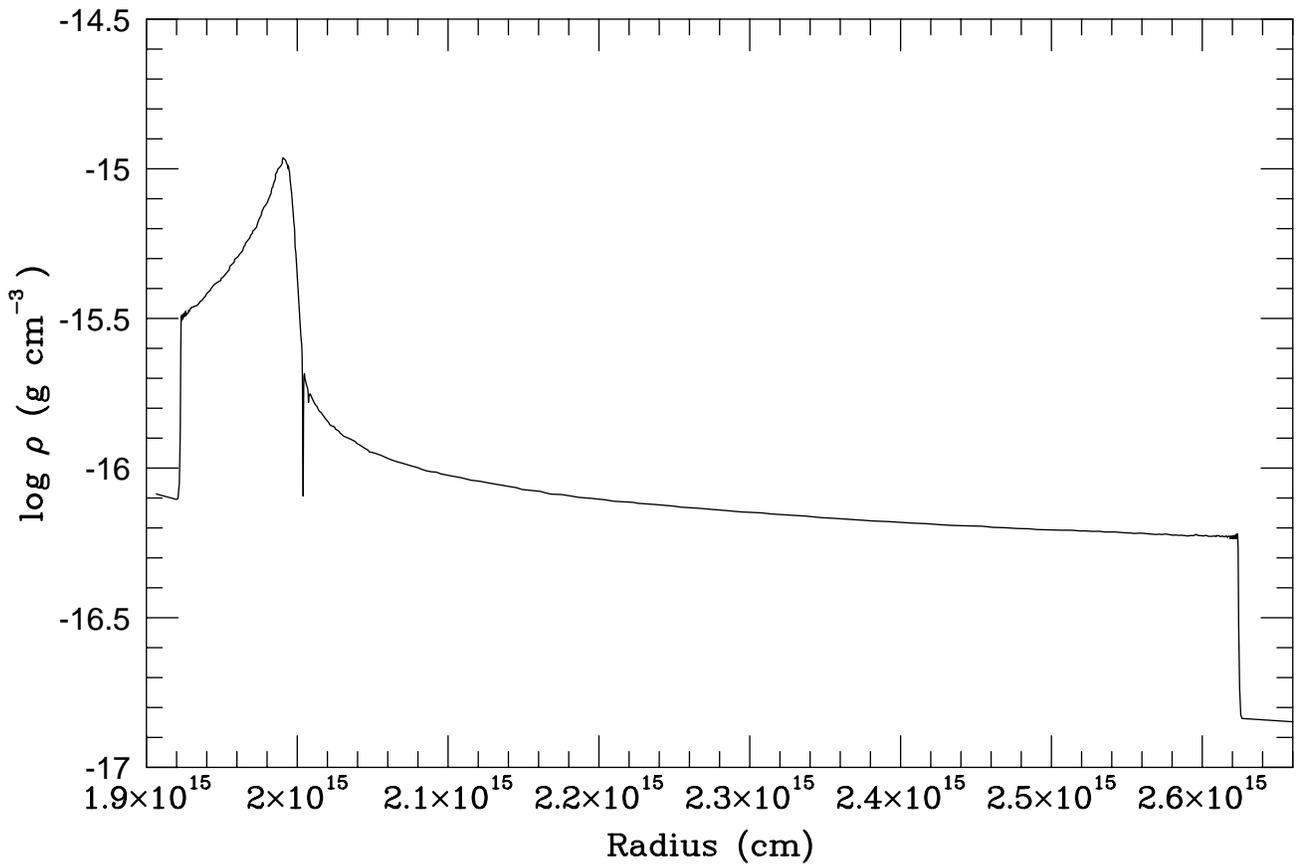

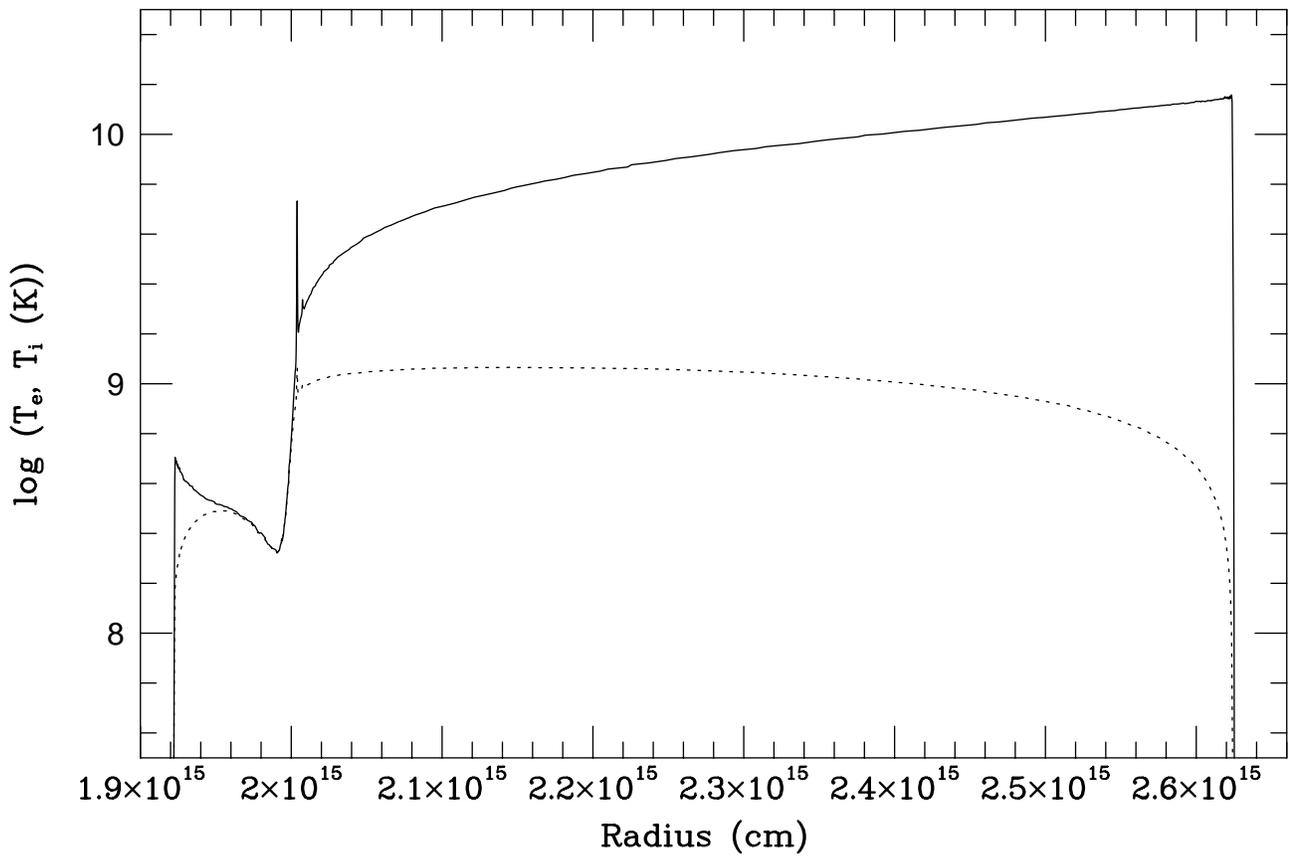

Fig. 4

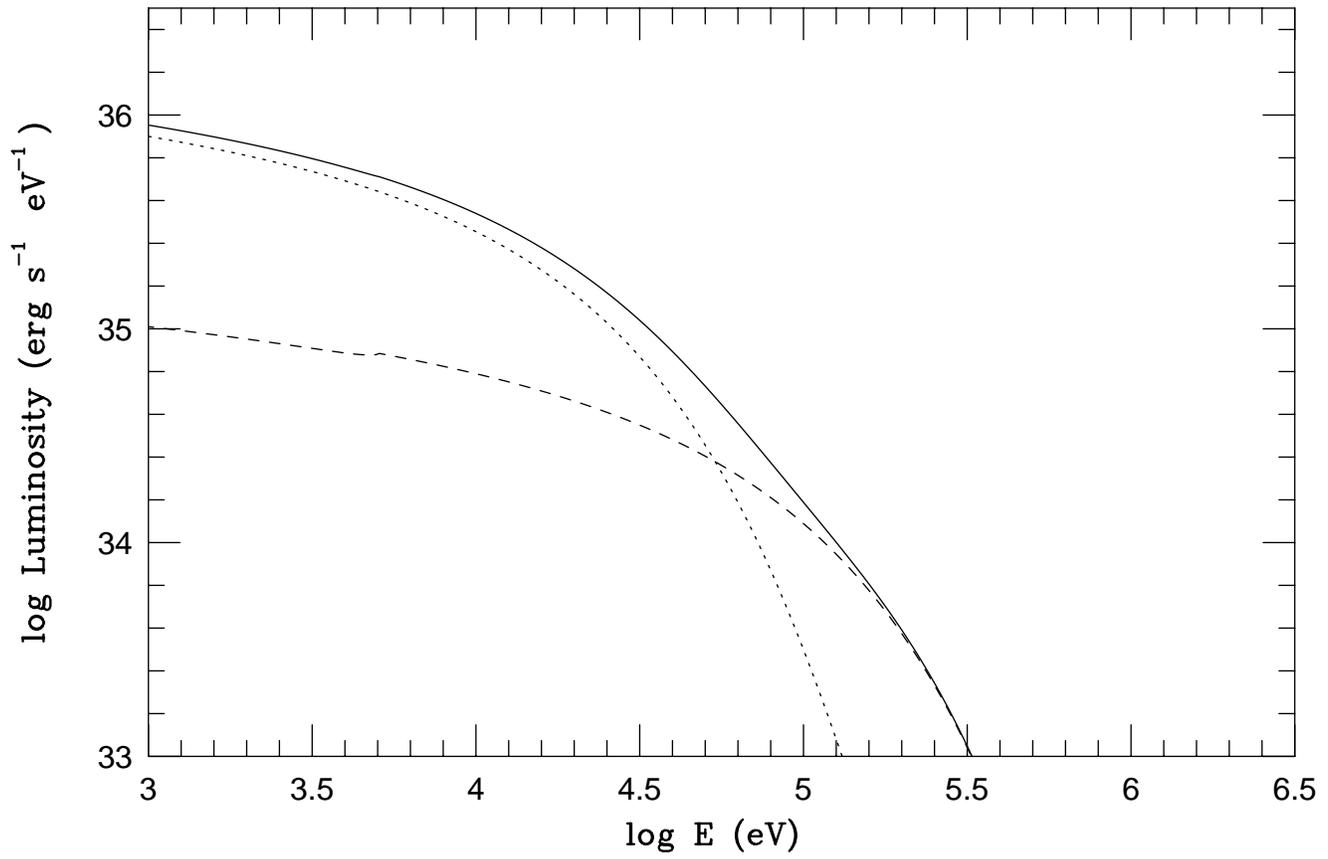

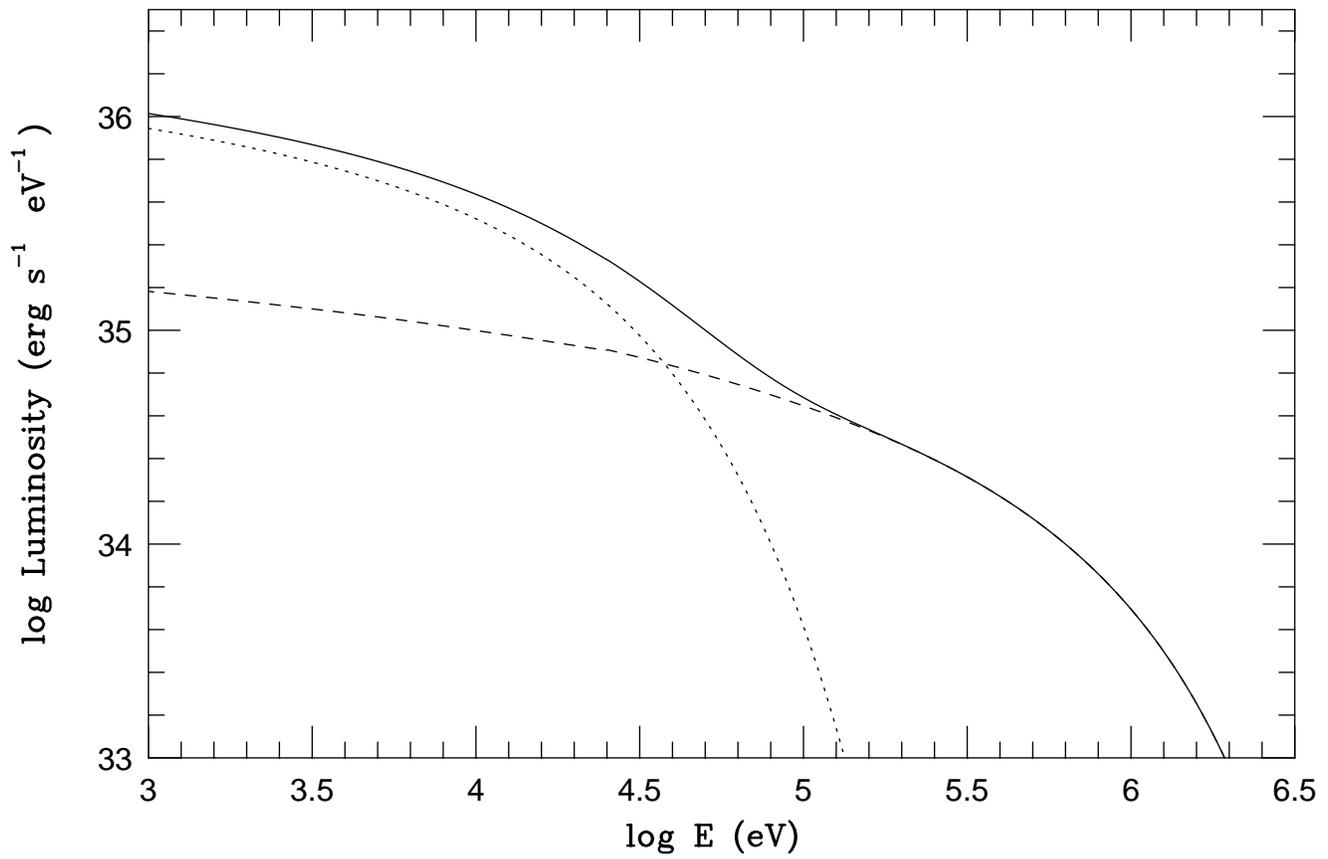

Fig. 5

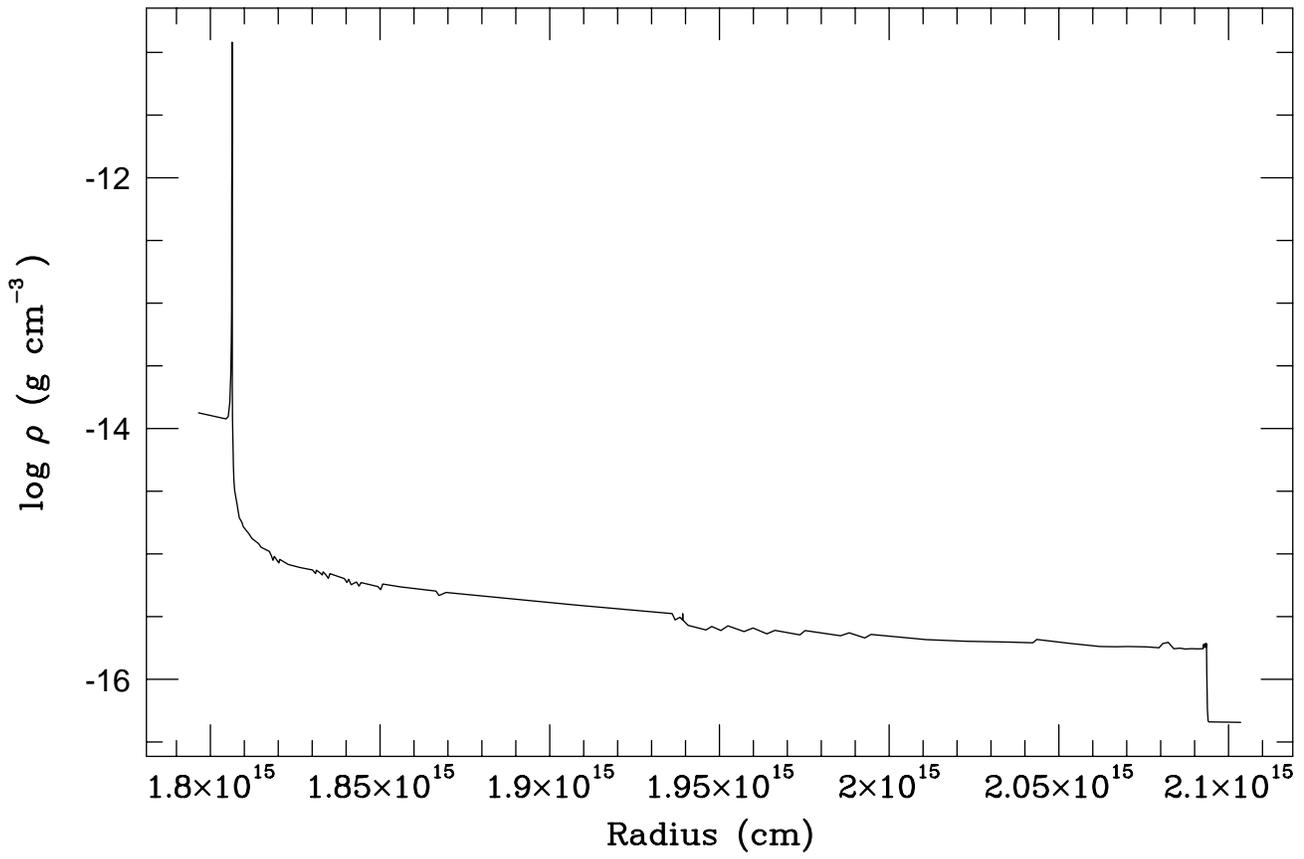

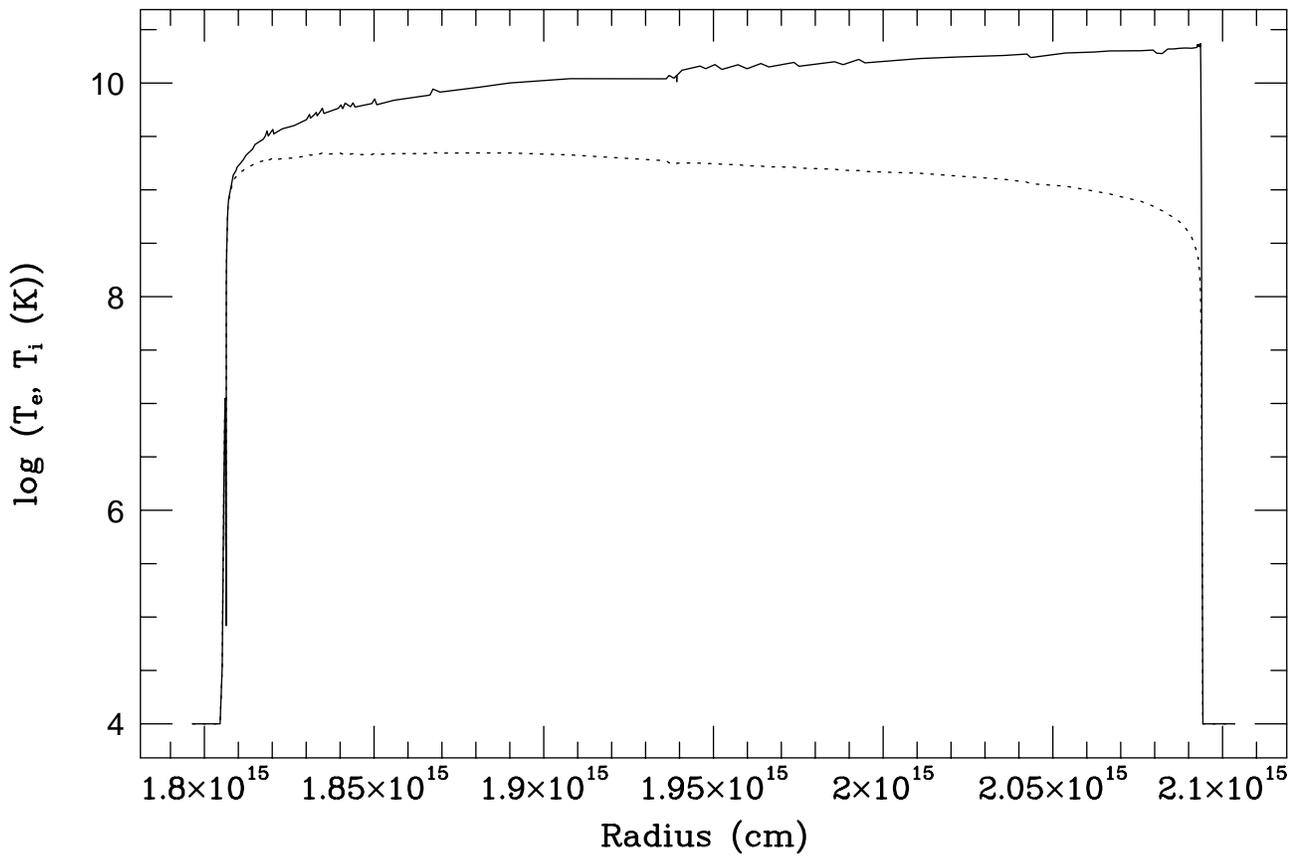

Fig. 6

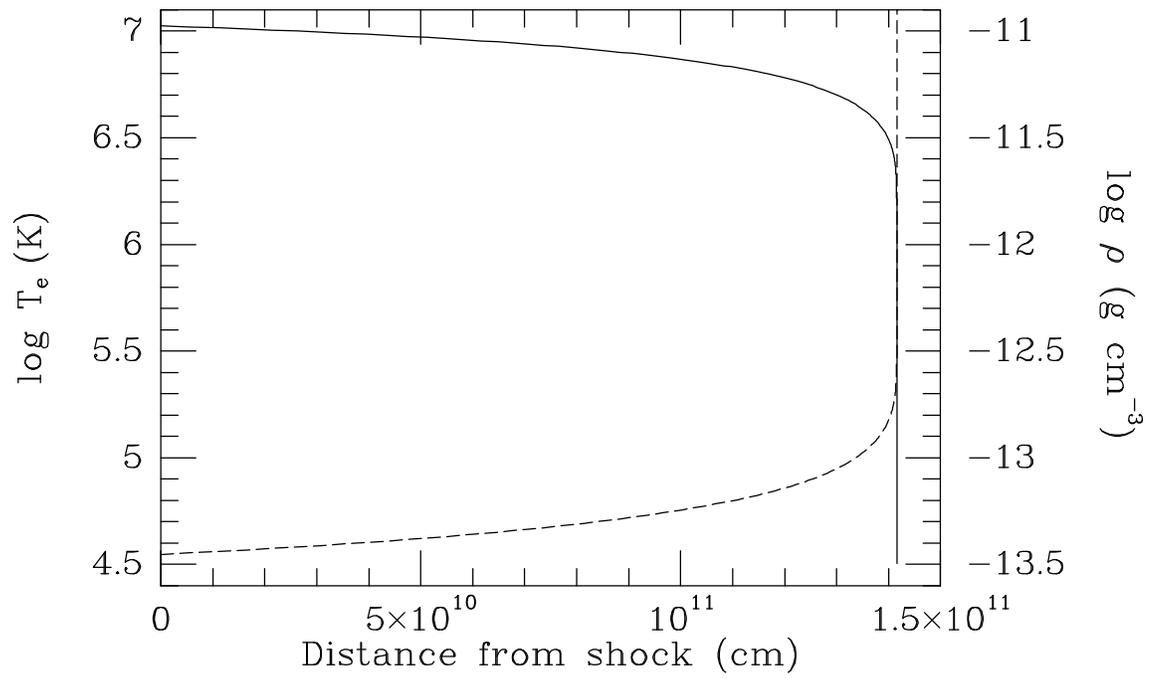

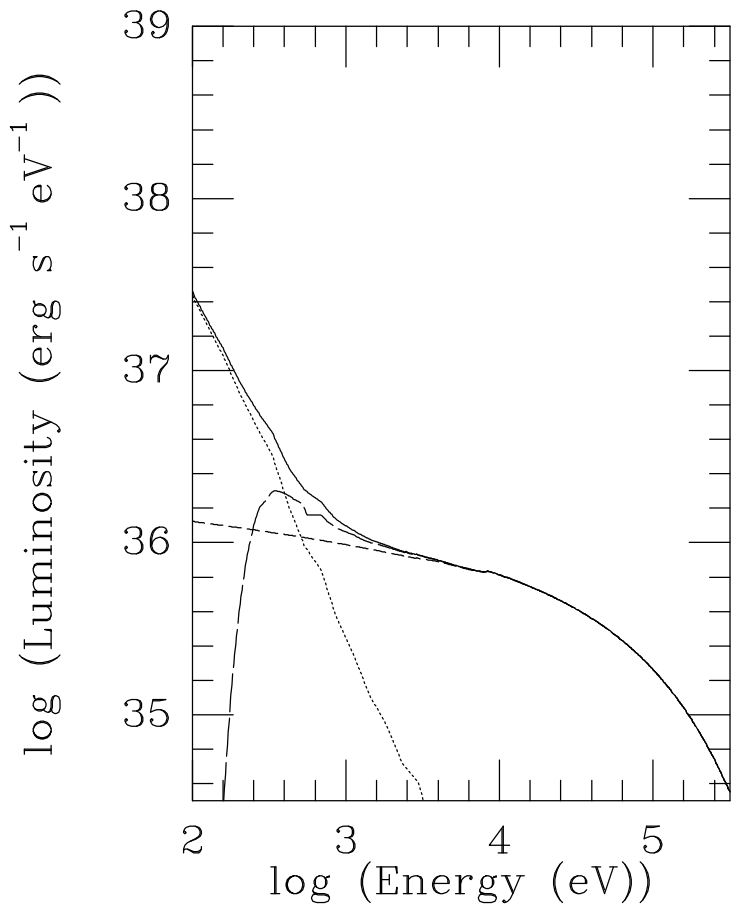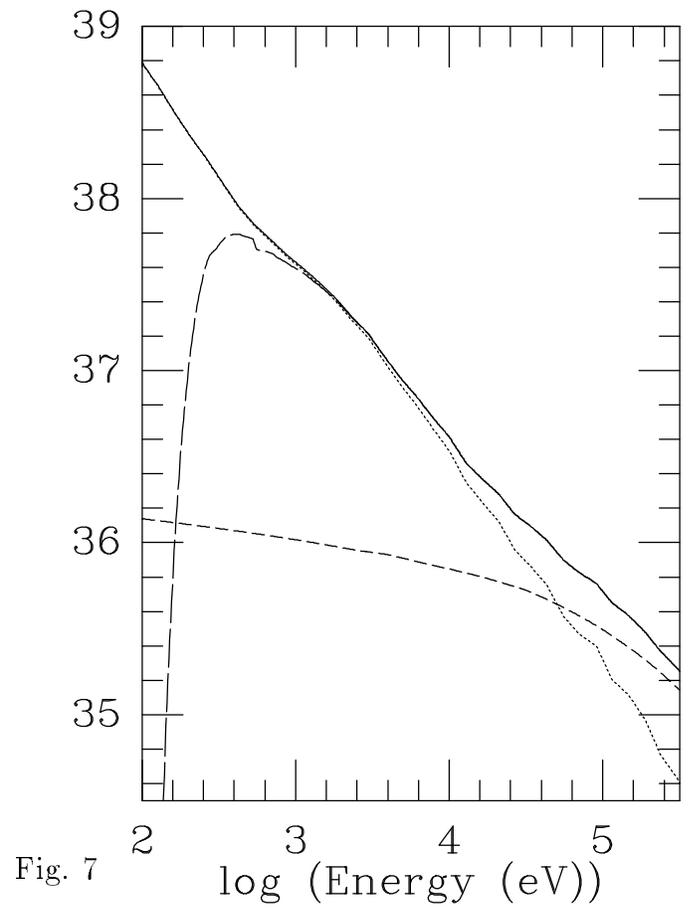

Fig. 7

Fig. 8

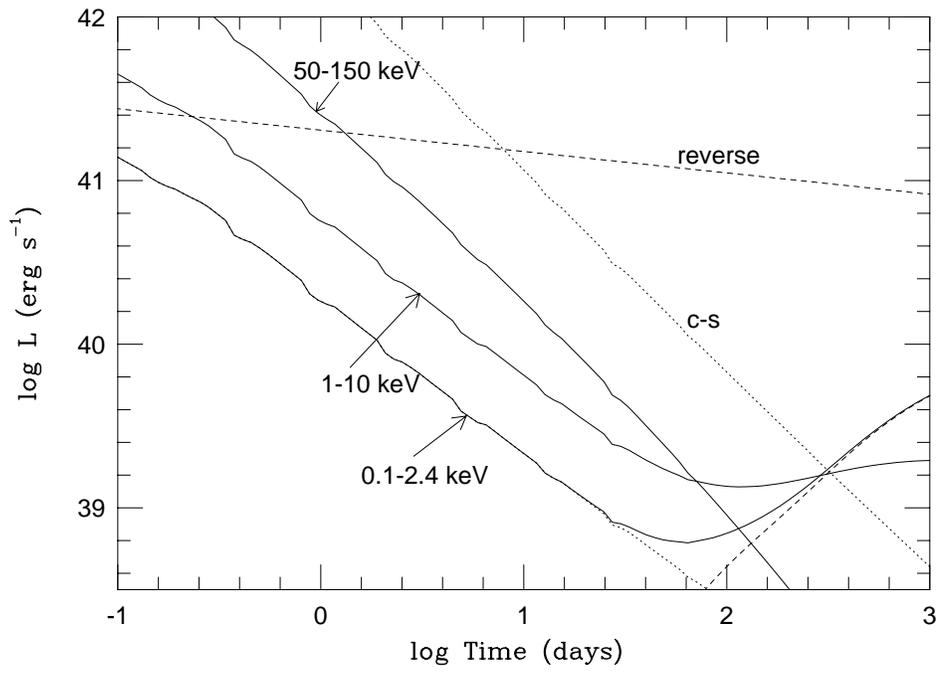

Fig. 9

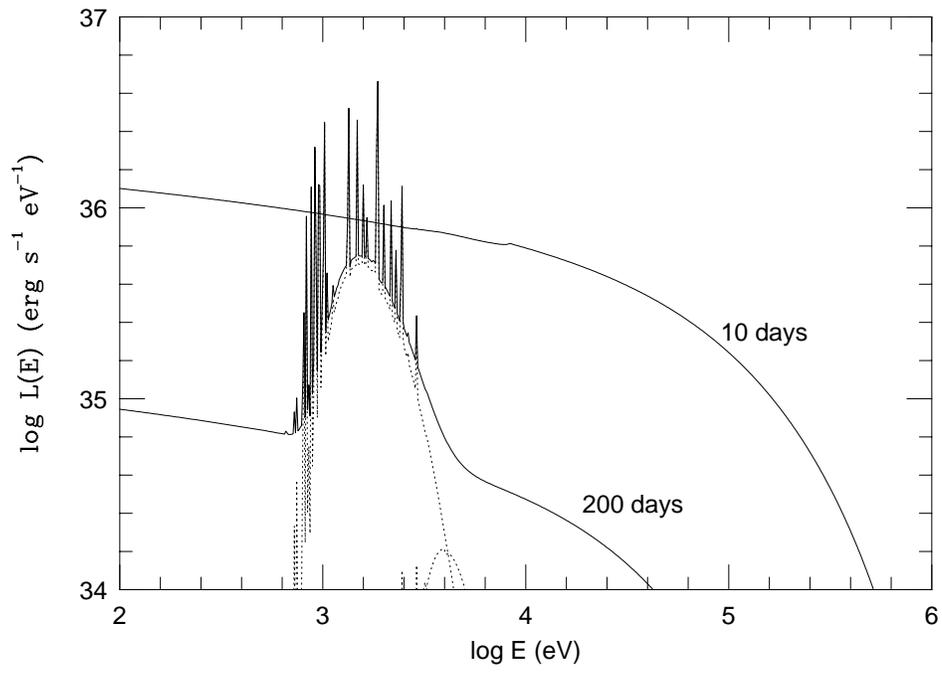

Fig. 10

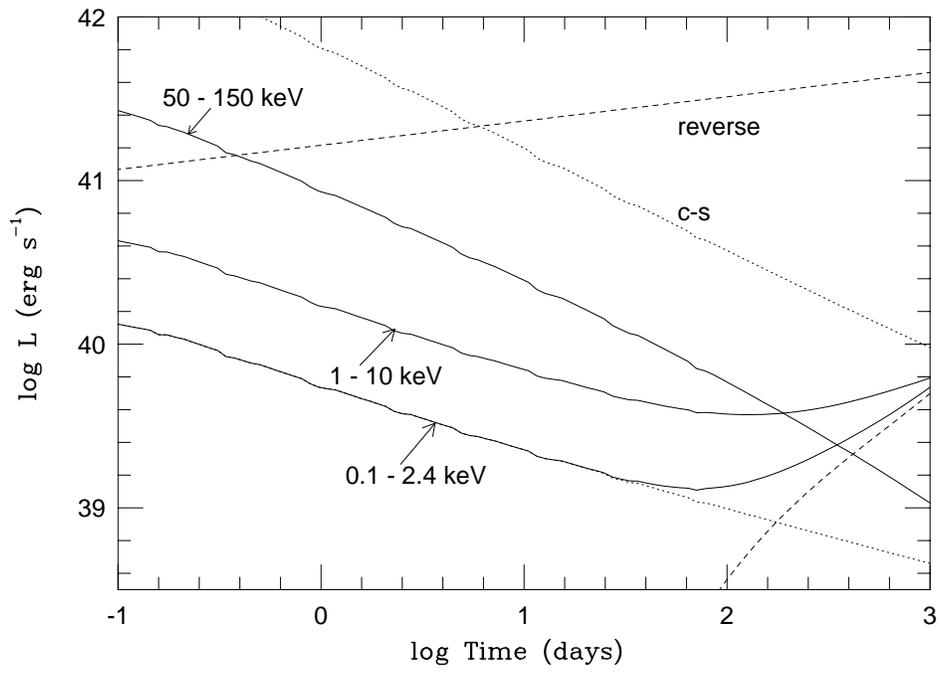

Fig. 11

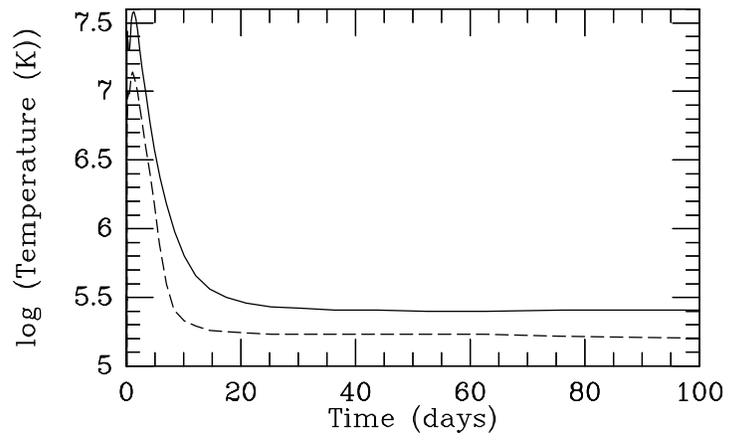

Fig. 12

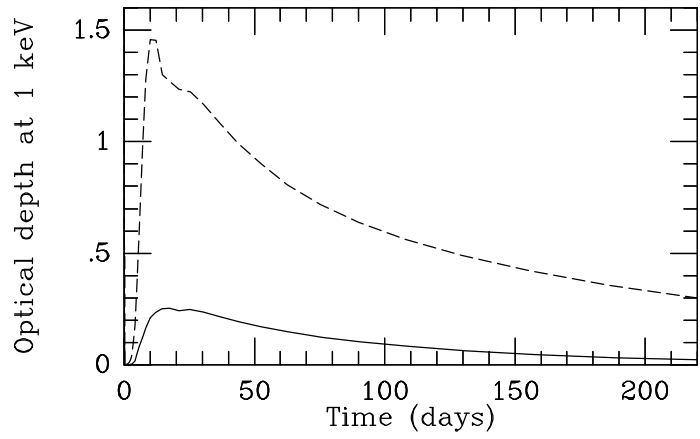